%% file: dijet.tex
\documentclass[oneside,12pt]{article}

\usepackage{hyperref}
\usepackage{graphicx}
\usepackage{array}
\usepackage{axodraw,feynarts}
\usepackage{color}
\usepackage[utf8]{inputenc}
\usepackage[square,comma,numbers,sort&compress]{natbib}
\usepackage[british]{babel}
\usepackage{amsmath,amssymb}

\makeatletter

\def\paragraph{\@startsection{paragraph}{4}{\z@}{+2.00ex plus
 +1ex minus +.2ex}{1.5ex plus .2ex}{\it\normalsize}}

\def\section{\@startsection {section}{1}{\z@}{+3.0ex plus +1ex minus
  +.2ex}{2.3ex plus .2ex}{\normalsize\bf\boldmath}}
\def\subsection{\@startsection{subsection}{2}{\z@}{+2.5ex plus +1ex
minus +.2ex}{1.5ex plus .2ex}{\normalsize\bf\boldmath}}
\def\subsubsection{\@startsection{subsubsection}{3}{\z@}{+3.25ex plus
 +1ex minus +.2ex}{1.5ex plus .2ex}{\normalsize\it}}

 \expandafter\ifx\csname mathrm\endcsname\relax\def\mathrm#1{{\rm #1}}\fi

\newcounter{saveeqn}

\@addtoreset{equation}{section}
\newcommand{\ie}{i.e.\ }
\newcommand{\eg}{e.g.\ }

\newcommand{\nnnl}{\nonumber\\} % No Number New Line

\newcommand{\ensuremathr}[1]{\ensuremath{\mathrm{#1}}}

\newcommand{\rd}{\ensuremathr{d}}
\newcommand{\ri}{\ensuremathr{i}}
\newcommand{\re}{\ensuremathr{e}}
\newcommand{\rw}{\ensuremathr{w}}

\newcommand{\rT}{\ensuremathr{T}}
\newcommand{\rR}{\ensuremathr{R}}
\newcommand{\rF}{\ensuremathr{F}}
\newcommand{\rL}{\ensuremathr{L}}

\newcommand{\PZ}{\ensuremathr{Z}}
\newcommand{\PW}{\ensuremathr{W}}

\newcommand{\PWpm}{\ensuremathr{W^\pm}}

\newcommand{\Pu}{\ensuremathr{u}}
\newcommand{\Pc}{\ensuremathr{c}}
\newcommand{\Pt}{\ensuremathr{t}}
\newcommand{\Pd}{\ensuremathr{d}}
\newcommand{\Ps}{\ensuremathr{s}}
\newcommand{\Pb}{\ensuremathr{b}}

\newcommand{\Pg}{\ensuremathr{g}}

\newcommand{\Pp}{\ensuremathr{p}}
\newcommand{\Pap}{\ensuremathr{\bar{p}}}

\newcommand{\Pj}{\ensuremath{j}} 

\newcommand{\jet}{\text{jet}}
\newcommand{\cut}{\text{cut}}

\newcommand{\sw}{\ensuremath{s_{\mathrm{w}}}}
\newcommand{\cw}{\ensuremath{c_{\mathrm{w}}}}
\newcommand{\alphaw}{\ensuremath{\alpha_{\mathrm{w}}}}
\newcommand{\alphas}{\ensuremath{\alpha_{\mathrm{s}}}}
\newcommand{\gs}{\ensuremath{g_{\mathrm{s}}}}
\newcommand{\gso}{\ensuremath{g_{\mathrm{s,0}}}}

\newcommand{\GeV}{\ensuremathr{GeV}}
\newcommand{\TeV}{\ensuremathr{TeV}}
\newcommand{\nb}{\ensuremathr{nb}}

\newcommand{\order}[1]{\ensuremath{ {\mathcal{O}\left( #1 \right)} }}

\DeclareMathOperator{\Deltauv}{\Delta_{\mathrm{\scriptscriptstyle UV}}}

\newcommand{\proc}{\ensuremath{\Pp\Pp\longrightarrow \Pj\Pj+X}}
\newcommand{\proctev}{\ensuremath{\Pp\Pap\longrightarrow \Pj\Pj+X}}

\newcommand{\sigtree}{\ensuremath{\sigma^0}}
\newcommand{\sigloop}{\ensuremath{\sigma^\mathrm{NLO}}}
\newcommand{\sigqcd}{\ensuremath{\sigma^0_\mathrm{QCD}}}

\newcommand{\deltree}{\ensuremath{\delta^\text{tree}_\mathrm{EW}}}
\newcommand{\delloop}{\ensuremath{\delta^\text{1-loop}_\mathrm{weak}}}
\newcommand{\delsum}{\ensuremath{\deltree{+}\delloop}}
\newcommand{\delsumabbr}{\ensuremath{\sum\delta}}

\newcommand{\shat}{\ensuremath{\hat{s}}}
\newcommand{\that}{\ensuremath{\hat{t}}}

\newcommand{\kt}{\ensuremath{k_{\rT}}}
\newcommand{\kti}{\ensuremath{k_{\rT,i}}}
\newcommand{\ktl}{\ensuremath{k_{\rT,1}}} % Leading
\newcommand{\ktsl}{\ensuremath{k_{\rT,2}}} % Sub-Leading
\newcommand{\ktcut}{\ensuremath{k_{\rT,\jet}^\cut}}
\newcommand{\mjj}{\ensuremath{M_{12}}}
\newcommand{\ystar}{\ensuremath{y^{*}}}
\newcommand{\ycut}{\ensuremath{y_\jet^\cut}}

\newcommand{\MSbar}{\ensuremath{\overline{\text{MS}}}}

\newcommand{\sigB}{\ensuremath{\sigma^\mathrm{B}}}
\newcommand{\sigV}{\ensuremath{\sigma^\mathrm{V}}}
\newcommand{\sigR}{\ensuremath{\sigma^\mathrm{R}}}
\newcommand{\sigC}{\ensuremath{\sigma^\mathrm{C}}}
\newcommand{\sigA}{\ensuremath{\sigma^\mathrm{A}}}
\newcommand{\CSV}{\ensuremath{\rd V_\text{dipole}}}
\newcommand{\CSI}{\ensuremath{\boldsymbol{I}(\epsilon)}}
\newcommand{\CSK}{\ensuremath{\boldsymbol{K}(x)}}
\newcommand{\CSP}{\ensuremath{\boldsymbol{P}(x,\mu_\rF^2)}}

\newcommand{\lsim}
{\mathrel{\raisebox{-.3em}{$\stackrel{\displaystyle <}{\sim}$}}}

\def\asymp#1%
{\mathrel{\raisebox{-.4em}{$\widetilde{\scriptstyle #1}$}}}

\def\Nequal#1%
{\mathrel{\raisebox{-.5em}{$\stackrel{=}{\scriptstyle\rm#1}$}}}
\newcommand{\dsl}[1]{\not \hspace{-0.7mm}#1}
\def\dsl{\mathpalette\make@slash}
\def\make@slash#1#2{\setbox\z@\hbox{$#1#2$}%
  \hbox to 0pt{\hss$#1/$\hss\kern-\wd0}\box0}

\def\myscale{0.85}

\def\draftdate{\relax}
\def\mda{\relax}
\def\mua{\relax}
\def\mla{\relax}
\def\Mda{\relax}
\def\Mua{\relax}
\def\Mla{\relax}
\def\draft{
\def\thtystars{******************************}
\def\sixtystars{\thtystars\thtystars}
\typeout{}
\typeout{\sixtystars**}
\typeout{* Draft mode!
         For final version remove \protect\draft\space in source file *}
\typeout{\sixtystars**}
\typeout{}
\def\draftdate{\today}
\def\mua{\marginpar[\boldmath\hfil$\uparrow$]%
                   {\boldmath$\uparrow$\hfil}%
                    \typeout{marginpar: $\uparrow$}\ignorespaces}
\def\mda{\marginpar[\boldmath\hfil$\downarrow$]%
                   {\boldmath$\downarrow$\hfil}%
                    \typeout{marginpar: $\downarrow$}\ignorespaces}
\def\mla{\marginpar[\boldmath\hfil$\rightarrow$]%
                   {\boldmath$\leftarrow $\hfil}%
                    \typeout{marginpar: $\leftrightarrow$}\ignorespaces}
\def\Mua{\marginpar[\boldmath\hfil$\Uparrow$]%
                   {\boldmath$\Uparrow$\hfil}%
                    \typeout{marginpar: $\uparrow$}\ignorespaces}
\def\Mda{\marginpar[\boldmath\hfil$\Downarrow$]%
                   {\boldmath$\Downarrow$\hfil}%
                    \typeout{marginpar: $\downarrow$}\ignorespaces}
\def\Mla{\marginpar[\boldmath\hfil$\Rightarrow$]%
                   {\boldmath$\Leftarrow $\hfil}%
                    \typeout{marginpar: $\leftrightarrow$}\ignorespaces}
\overfullrule 5pt
\oddsidemargin -15mm
\oddsidemargin -10mm
\marginparwidth 29mm
}

\begin{document}

\thispagestyle{empty}
\def\thefootnote{\fnsymbol{footnote}}
\setcounter{footnote}{1}
\null
\mbox{}\hfill  
FR-PHENO-2012-024 \\
\vskip 0cm
\vfill
\begin{center}
  {\Large \boldmath{\bf 
      Weak radiative corrections to\\[.5em]
      dijet production at hadron colliders
    }
    \par} \vskip 2.5em
  {\large
    {\sc Stefan Dittmaier, Alexander Huss \\[.3em]
      and Christian Speckner
    }\\[1ex]
    {\normalsize \it 
      Albert-Ludwigs-Universit\"at Freiburg, Physikalisches Institut, \\
      D-79104 Freiburg, Germany
    }
    \\[2ex]
  }
  \par \vskip 1em
\end{center}\par
\vskip .0cm \vfill {\bf Abstract:} \par
We present the calculation of the most important electroweak
corrections to dijet production at the LHC and the Tevatron,
comprising tree-level effects of $\order{\alphas\alpha,\,\alpha^2}$ 
and weak loop corrections of $\order{\alphas^2\alpha}$.
Although negligible for integrated cross sections, these
corrections can reach $10{-}20\%$ in the TeV range for transverse
jet momenta $\kt$. 
Our detailed discussion of numerical results comprises distributions
in the dijet invariant mass and in the transverse momenta of the
leading and subleading jets.
We find that the weak loop corrections amount to
about $-12\%$ and $-10\%$ for leading jets with $\kt\sim3~\TeV$ at the
$14~\TeV$ LHC and $\kt\sim800~\GeV$ at the Tevatron, respectively.
The electroweak tree-level contributions are of the same generic
size and typically positive at the LHC and negative at the Tevatron
at high energy scales.
Generally the corrections to the dijet invariant mass distributions
are smaller by at least a factor of two as compared to the
corresponding reach in the $\kt$ distributions,
because unlike the $\kt$ spectra the invariant-mass distributions
are not dominated by the Sudakov regime at high energy scales.
\par
\vskip 1cm
\noindent
October 2012
\par
\null
\setcounter{page}{0}
\clearpage
\def\thefootnote{\arabic{footnote}}
\setcounter{footnote}{0}

\section{Introduction}
\label{sec:intro}

The unprecedented energy regime that is accessible at the LHC
allows for the investigation of the laws of physics at the smallest
distances. 
The inclusive production of two jets (dijets) at the LHC, $\proc$, 
allows for a detailed study of QCD at TeV energies. 
Furthermore, several extensions of the Standard Model predict new
heavy particles which might be visible via dijet signatures in the
detector~\cite{Harris2011}.  
Some examples are excited states of composite quarks
$q^{*}$, string resonances, new heavy gauge bosons $\PW^{\prime},
\PZ^{\prime}$, etc. 
Inclusive jet and dijet production has been analyzed by the
ATLAS~\cite{Aad:2011fc} 
and CMS~\cite{CMS-jets} 
collaborations at a centre-of-mass (CM) energy of $7~\TeV$
giving sensitivity to dijet invariant masses of up
to $5~\TeV$ and jet transverse momenta of up to $2~\TeV$
at the LHC.
The Tevatron experiments CDF~\cite{Aaltonen:2008eq}
and D0~\cite{Abazov:2011vi} have investigated jet production up to
transverse momenta of several hundreds of GeV.
At the current level of experimental and theoretical accuracy,
the SM is able to describe data quite well.
At the LHC design CM energy of $14~\TeV$, the energy reach 
will even go deeper into the TeV range, so that
theoretical predictions especially have to carefully include
radiative corrections that are sensitive to high scales.

The results for the production of two jets at leading-order (LO)
accuracy in QCD had been available 
\cite{Combridge1977} 
long before higher-order corrections were established.
Later, in the 1990s,
the differential cross sections to inclusive single-jet and two-jet
production were discussed at next-to-leading order (NLO) accuracy in QCD
\cite{Ellis1990,Ellis1992,Giele1994a}.
Currently, enormous effort is put into the calculation of the NNLO
QCD corrections to dijet production (see \eg Refs.~\cite{%
Gehrmann:2010rj,GehrmannDeRidder:2010zz,Bolzoni:2010bt,%
Boughezal:2010mc,GehrmannDeRidder:2011aa, Gehrmann2011,Ridder:2012ja}
and references therein).
The purely weak
corrections of $\order{\alphas^2\alpha}$ have been calculated for the
single-jet-inclusive cross section in Ref.~\cite{Moretti2006c}, 
and preliminary results of the weak corrections to the dijet production
were published in Ref.~\cite{Scharf2009}. 
The two results, however, do not seem compatible with each other.
Electroweak corrections also were calculated for the related
process of bottom-jet production~\cite{Kuhn2010}. 

In spite of their suppression
by the small value of the coupling constant
$\alpha$, the electroweak (EW) corrections can become large in the
high-energy domain~\cite{Fadin2000a, Ciafaloni2000a, Hori2000,
  Melles2002, Beenakker2002, Denner2003b, Jantzen2005, Jantzen2005a}. 
This is due to the appearance of Sudakov-type and other high-energy
logarithms that result from the virtual exchange of soft or collinear
massive weak gauge bosons.
The leading term is given by
$\alphaw\ln^2\left(Q^2/M_\PW^2\right)$, 
where $Q$ denotes a typical energy scale of the hard-scattering reaction,
$M_\PW$ is the W-boson mass, and 
$\alphaw=\alpha/\sw^2=e^2/(4\pi\sw^2)$ is derived from the $SU(2)$
gauge coupling $e/\sw$ with $\sw$ denoting the sine of the weak mixing
angle $\theta_\rw$.
In the case of massless gauge bosons, \eg in QED or QCD, these
logarithms are connected to the well-known infrared
divergences and are cancelled against the corresponding
real-emission corrections.
For the massive gauge bosons $\PW$ and $\PZ$, no
such singularities occur, since their masses provide a physical
cut-off and the additional radiation of real $\PW$ or $\PZ$ bosons can
be experimentally reconstructed to a large extent, so that $\PW$/$\PZ$
bremsstrahlung corresponds to a different class of processes. 
Thus, at high scales $\lvert Q^2\rvert\gg M_\PW^2$, which are accessible
at the LHC and the Tevatron, 
the above Sudakov-type logarithms can produce large
negative corrections, as only some fractions are compensated by
unresolved $\PW$/$\PZ$ emission~\cite{Baur2007}. 
It turns out that large compensations can occur between different
electroweak logarithms~\cite{Kuhn2000a,Kuhn2000}, so that a full
fixed-order calculation desirable.

As it will be discussed in more detail below, a gauge-invariant
classification of the EW corrections into photonic and purely
weak corrections is feasible for our process.
Guided by the logarithmic enhancements, 
we restrict ourselves to the calculation of the purely weak
corrections in this paper. 
The calculation can be complemented by the photonic corrections
at a later time to produce results for the full EW
corrections at the order $\alphas^2\alpha$. 

The paper is organized as follows: 
In Section~\ref{sec:dijet} we set up our conventions
(Sect.~\ref{sec:setup}) and discuss our strategy for the calculation of
the NLO corrections (Sect.~\ref{sec:nlo}).
The numerical results are presented in Section~\ref{sec:num}, which
comprises integrated cross sections as well as differential
distributions and the comparison to other work.
Section~\ref{sec:concl} contains our conclusions.

\section{Dijet production in hadronic collisions}
\label{sec:dijet}

\subsection{Conventions and calculational setup}
\label{sec:setup}

We consider the hadronic process
\begin{equation}
  \label{eq:10}
  A(p_A) + B(p_B) \rightarrow \Pj(k_1) + \Pj(k_2) + X,
\end{equation}
where the assignment of the four-momenta to the respective particles
is indicated in parentheses.
We further assume that the momenta
$k_1$ and $k_2$ are sorted in a descending order with respect to their
transverse momenta, \ie $\ktl \ge \ktsl$, referring to the
associated jets as the leading and subleading jet, respectively.
The hadronic cross section is given by the incoherent sum over the
different partonic subprocesses that contribute to the final
state under consideration, convoluted with the respective parton
distribution functions (PDFs),
\begin{equation}
  \label{eq:11}
  \sigma_{AB}(p_A,p_B) 
  = \sum_{a,b} \int_0^1\rd x_a \int_0^1\rd x_b 
  \;f_{a\vert A}(x_a,\mu_\rF^2) f_{b\vert B}(x_b,\mu_\rF^2)
  \hat{\sigma}_{ab}(p_a,p_b).
\end{equation}
The PDF $f_{a\vert A}(x_a,\mu_\rF^2)$ plays the role of a generalized 
number density to find a parton $a$ carrying the momentum fraction
$x_a$ of the parent hadron $A$ with $p_{a}=x_{a} p_{A}$ denoting the
four-momentum of the incoming parton to the hard scattering. 
We work in the QCD-improved parton model using the five-flavour scheme
with $N_f=5$ massless quarks $q=\Pu,\Pd,\Pc,\Ps,\Pb$. 
The partonic subprocesses that contribute to the above scattering
reaction at NLO can be generically written as
\begin{equation}
  \label{eq:13}
  a(p_a) + b(p_b) \rightarrow c(k_c) + d(k_d) \left( +e(k_e) \right),
\end{equation}
where $a,b,c,d,e\in\lbrace g,\Pu,\Pd,\Pc,\Ps,\Pb \rbrace$
if only weak corrections are considered.
For photonic corrections also the photon has to be included as
a possible external particle state.
The additional emission of parton $e$ appears in the real NLO
correction to this process. 
The momenta $k_1$ and $k_2$ in Eq.~\eqref{eq:10} emerge from the
recombination procedure of the jet algorithm. 
At LO, however, we simply have $k_1=k_c,\;k_2=k_d$, with
$\ktl=\ktsl$.
Owing to the mass degeneracy ($m_q=0$) of the external quarks and the
unitarity of the CKM matrix, the effect of the non-diagonal CKM structure
vanishes in most contributions after taking the flavour sums. 
The only exception is the case of a $\PW$-boson exchange in the $s$-channel,
where the different weights from the PDFs spoils the cancellation. 
However, this dependence turns out to be negligible
and we set the CKM matrix to unity in our calculation.

\begin{figure}
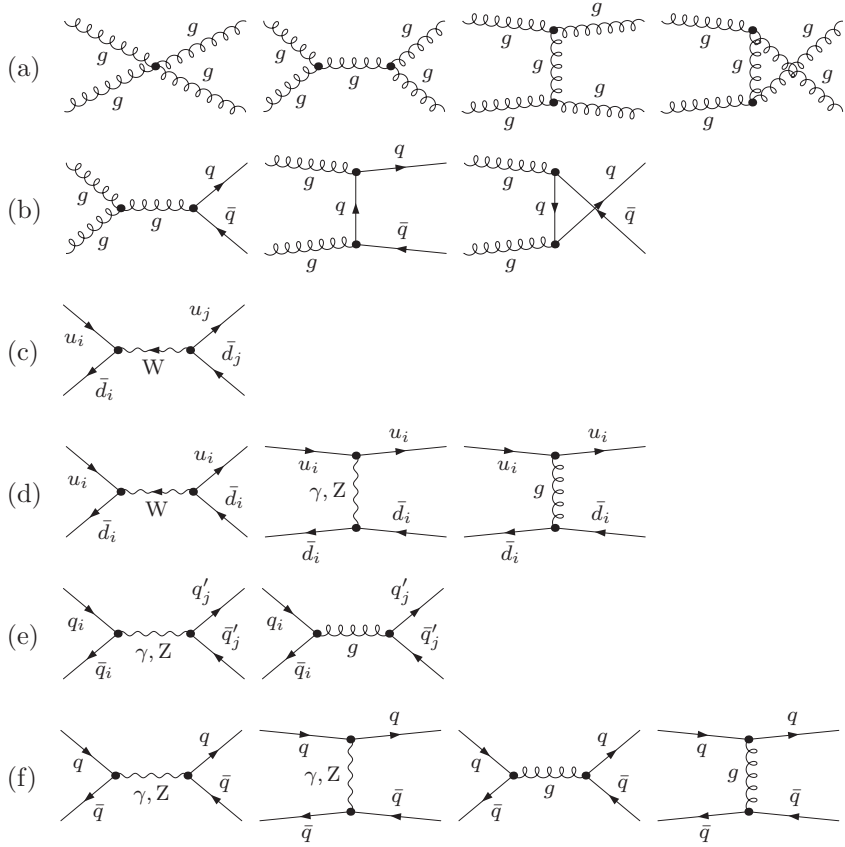

  \centering
  \unitlength=0.75bp
  \scriptsize
  \begin{minipage}{430\unitlength}
      \raisebox{30\unitlength}{\footnotesize(a)}~
  \raisebox{-15\unitlength}[70\unitlength][0mm]{\input{feyndiags/Born4G}}
\\
      \raisebox{30\unitlength}{\footnotesize(b)}~
  \raisebox{-15\unitlength}[70\unitlength][0mm]{\input{feyndiags/Born2G2Q}}
\\
      \raisebox{30\unitlength}{\footnotesize(c)}~
  \raisebox{-15\unitlength}[70\unitlength][0mm]{\input{feyndiags/Born4Q1}}
\\
      \raisebox{30\unitlength}{\footnotesize(d)}~
  \raisebox{-15\unitlength}[70\unitlength][0mm]{\input{feyndiags/Born4Q2}}
\\
      \raisebox{30\unitlength}{\footnotesize(e)}~
  \raisebox{-15\unitlength}[70\unitlength][0mm]{\input{feyndiags/Born4Q3}}
\\
      \raisebox{30\unitlength}{\footnotesize(f)}~
  \raisebox{-15\unitlength}[70\unitlength][0mm]{\input{feyndiags/Born4Q4}}

  \end{minipage}
  \caption{Tree-level Feynman graphs (a--f) to the processes
    (\ref{eq:subproc}a--f), respectively.}
  \label{fig:lodiags}
\end{figure}

The partonic subprocesses can be classified as follows:
\begin{subequations}
  \label{eq:subproc}
  \begin{align}
    \label{eq:4g}
    &g+g\rightarrow g+g,\\
    \label{eq:2g2q}
    &g+g\rightarrow q+\bar{q},\\
    \label{eq:uddg}
    &u_i+\bar{d}_i\rightarrow u_j+\bar{d}_j, \quad(i\ne j),\\
    \label{eq:udsg}
    &u_i+\bar{d}_i\rightarrow u_i+\bar{d}_i,\\
    \label{eq:qqdg}
    &q_i+\bar{q}_i\rightarrow q^\prime_j+\bar{q}^\prime_j,
    \quad(i\ne j),\\
    \label{eq:qqsg}
    &q+\bar{q}\rightarrow q+\bar{q},
  \end{align}
\end{subequations}
where $i$, $j$ denote the generation indices.
We note that the processes listed in Eq.~\eqref{eq:subproc} are merely
representatives of a class that also include the reactions that are
related by crossing symmetry.
First, we categorize the processes according to the number of gluons
and quarks that appear as external particles into the four-gluon (a),
the two-gluon--two-quark (b), and the four-quark (c--f) processes. 
The four-quark processes can be subdivided into processes
that involve a $\PW$-boson exchange diagram (c,d) and those that only
contain neutral-boson exchange diagrams (e,f).
A further distinction is made by distinguishing processes that involve
both $s$-channel and $t$-channel diagrams (c,e), and those that only
include either one (d,f). 
The LO Feynman graphs to the above process classes are shown in
Fig.~\ref{fig:lodiags}. 
Furthermore, we can exploit the symmetry of the matrix element with
respect to the interchange of the generation index of the first two
quark generations.
This reduces the number of independent amplitudes that need to be
evaluated and speeds up the numerical evaluation. 

The electroweak coupling constant is derived from the Fermi constant in
the $G_\mu$ scheme via the following relation
\begin{equation}
  \label{eq:6}
  \alpha_{G_\mu} = \frac{\sqrt{2}}{\pi}G_\mu M_\PW^2
  \left(1-\frac{M_\PW^2}{M_\PZ^2}\right).
\end{equation}
This input-parameter scheme
avoids large logarithms of the light fermion masses generated by the
running of the coupling constant $\alpha(Q)$ from the Thomson limit ($Q=0$) to the
electroweak scale ($Q\sim M_\PW$) and furthermore absorbs universal corrections
induced by the $\rho$ parameter (see \eg Ref.~\cite{Dittmaier2002}).

In order to describe the resonances of the intermediate vector bosons
$\PZ$ and $\PW$, we employ the complex-mass scheme~\cite{Denner1999,Denner2005a},
which fully respects gauge invariance. 
In this approach the square of the gauge-boson mass is defined as
the position of the pole in the complex $k^2$ plane of the respective
propagator with momentum $k$.
The consistent replacement of the (squared) gauge-boson masses by complex values,
\begin{equation}
  \label{eq:1}
  M_V^2\rightarrow\mu_V^2 = M_V^2 - \ri M_V \Gamma_V, \quad V=\PW,\PZ,
\end{equation}
induces the adaption of all real quantities.
In particular, this results in a complex weak-mixing angle
$\theta_\rw$: 
\begin{equation}
  \label{eq:2}
  \cos^2\theta_\rw \equiv \cw^2 = \frac{\mu_\PW^2}{\mu_\PZ^2}, \quad
  \sin^2\theta_\rw \equiv \sw^2 = 1-\cw^2.
\end{equation}

In order to ensure the correctness of the presented results two
independent calculations have been performed, resulting in two separate
implementations for the numerical evaluation.
Both calculations employ the Feynman-diagrammatic approach in the
't~Hooft--Feynman gauge for the loops
and the Catani--Seymour dipole subtraction 
approach~\cite{Catani1997,Catani2002} to isolate and cancel infrared (IR) divergences.
The results of the two calculations are in mutual agreement.

In the first calculation all tree-level amplitudes are calculated and
implemented by hand using the Weyl--van-der-Waerden spinor formalism
as worked out in Ref.~\cite{Dittmaier1999}.
The virtual corrections are calculated using the \textsc{Mathematica}
Packages \textsc{FeynArts}~$3.6$~\cite{Hahn2001} and
\textsc{FormCalc}~$6.2$~\cite{Hahn1999}.
The one-loop integrals are evaluated using a modified version of the
\textsc{LoopTools}~$2.4$~\cite{Hahn1999} library, which was
supplemented by the loop integrals with dimensionally
regularized IR divergences that were not included in version
$2.4$.
Additionally, an interface is implemented between \textsc{LoopTools}
and the \textsc{Collier} library, which is based on the
results of Refs.~\cite{Denner2006,Denner:2010tr} for tensor and scalar
one-loop integrals, respectively.  
This allows to utilize the
unmodified code generated by \textsc{FormCalc}, while resorting to
\textsc{Collier} for the evaluation of the loop integrals. 
The results of the two approaches are in perfect mutual agreement.
The numerical integration is performed using the adaptive
Monte Carlo algorithm \textsc{Vegas}~\cite{Lepage}, where a specific
phase-space parametrization is chosen. 

The second calculation uses \textsc{FeynArts} $1.0$~\cite{Kublbeck1990} for
generating the tree-level and one-loop diagrams and in-house \textsc{Mathematica}
routines to obtain an analytic result which then is exported as \textsc{Fortran}
source code. The loop integrals are evaluated using the \textsc{Collier} loop
library. 
The finite Catani--Seymour
dipole subtraction terms and real emission matrix elements are built
around amplitudes generated with the \textsc{O'Mega}~\cite{Moretti:2001zz}
matrix-element generator. Adaptive single-channel Monte Carlo integration is
implemented using the \textsc{Vamp}~\cite{Ohl:1998jn} library. Contrary to the
first calculation, the second calculation does not implement subprocesses
involving external bottom quarks which only amount to $\approx3\%$ of the
LO cross section and thus can be safely neglected at order $\alphas^2\alphas$.
For all other subprocesses, both calculations are in excellent agreement for both
the integrated and differential cross sections.

\subsection{Structure of the NLO calculation}
\label{sec:nlo}

The calculation of the LO cross section is based on the full SM, \ie
all vector bosons, including the photon $\gamma$, are included. 
The Born diagrams for each process class defined in the previous
section are shown in Fig.~\ref{fig:lodiags}.

At NLO we classify the corrections into photonic and purely
weak corrections.
This is possible, because each diagram in the full NLO correction, defined
by the order $\alphas^2\alpha$, contains exactly one electroweak gauge
boson due to the single power in $\alpha$. 
We can therefore uniquely assign each NLO contribution to either a
photonic or a purely weak correction. 
The photonic corrections would constitute
the $\order{\alphas^2\alpha}$ corrections in a hypothetical gauge
theory with the group $SU(3)_\mathrm{C}\times U(1)_\mathrm{QED}$. 
Therefore, they form a gauge-invariant subset and consequently so do
the remaining purely weak corrections. 
In the following, we will use the notation $\order{\alphas^2\alphaw}$ to refer to
the purely weak contributions. 

This gauge-invariant classification of the corrections allows us to
tackle each class successively. 
In this work we present the calculation of the
$\order{\alphas^2\alphaw}$ corrections, as motivated in the
introduction. 
The inclusion of the photonic corrections in order to obtain the full
EW corrections at the order $\alphas^2\alpha$ is left to the future.

It is known~\cite{Bell2010,Baur2007} that partial cancellations can
occur between the virtual weak corrections and the real emission of
massive gauge bosons. 
However, to which extent this compensation takes place strongly
depends on the experimental setup to reconstruct $\PW$ and $\PZ$ bosons.
Considering that the weak-boson emission is a tree-level process which
can be easily simulated with fully LO automatic tools, 
they are not further considered here.

\subsubsection{Virtual corrections}
\label{sec:virt}

\begin{figure}
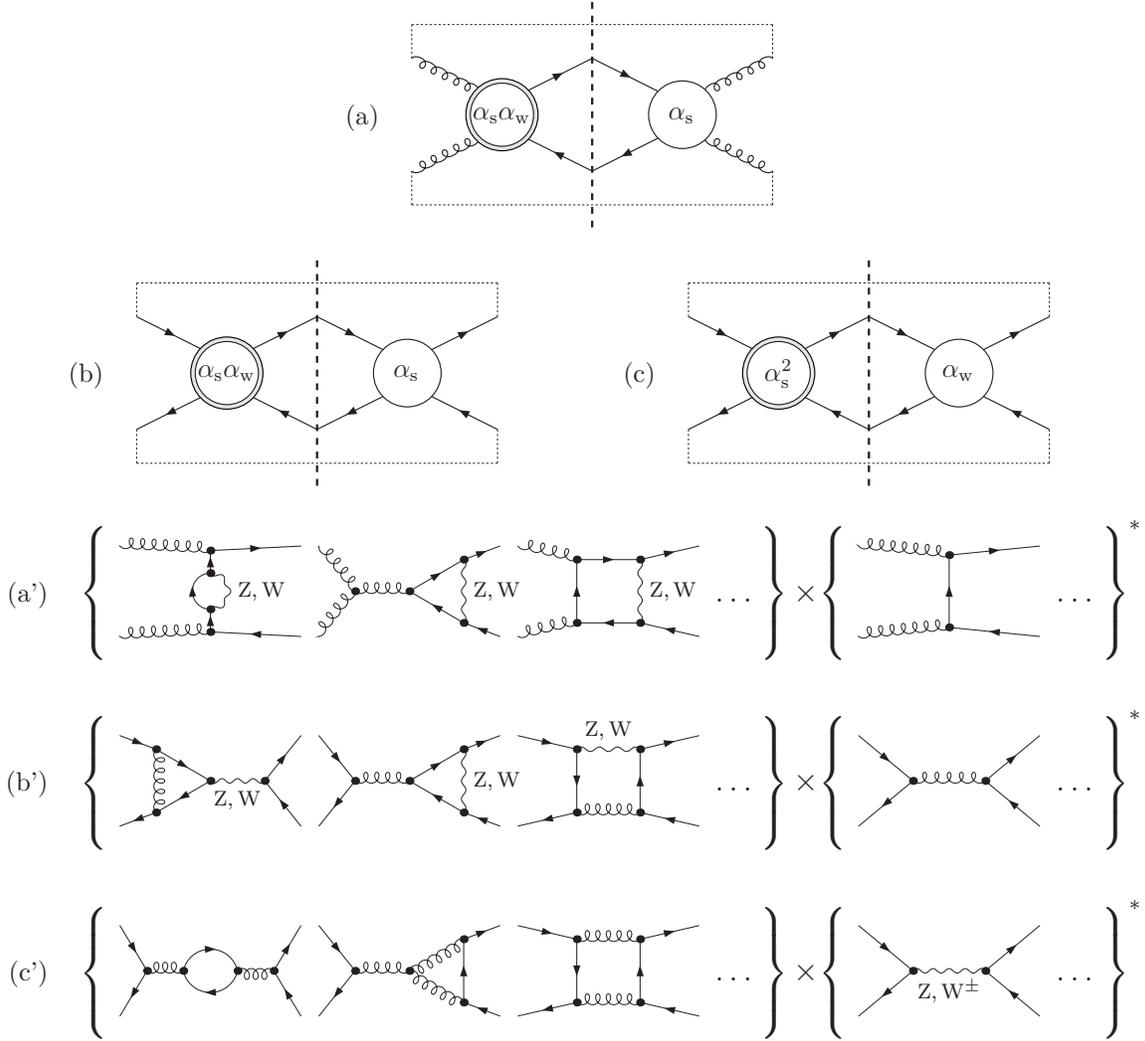

  \centering
  \raisebox{1.38cm}{\footnotesize(a)}~\scalebox{\myscale}{\input{feyndiags/IntVg}}\\[1em]
  \raisebox{1.38cm}{\footnotesize(b)}~\scalebox{\myscale}{\input{feyndiags/IntV1}}
  \hspace{1.0cm}
  \raisebox{1.38cm}{\footnotesize(c)}~\scalebox{\myscale}{\input{feyndiags/IntV2}}
  {\unitlength=0.75bp
  \begin{align*}
    \raisebox{30\unitlength}{\text{\footnotesize(a')}\quad
      $\left\lbrace\vphantom{\rule{0mm}{40\unitlength}}\right.$}
    {\scriptsize
  \raisebox{-15\unitlength}[70\unitlength][0mm]{\input{feyndiags/VGSW}}
}
    \raisebox{30\unitlength}{$\ldots\left.\vphantom{\rule{0mm}{40\unitlength}}\right\rbrace\times$}
    &\raisebox{30\unitlength}{$\left\lbrace\vphantom{\rule{0mm}{40\unitlength}}\right.$}
    {\scriptsize
  \raisebox{-15\unitlength}[70\unitlength][0mm]{\input{feyndiags/BGS}}
}
    \raisebox{30\unitlength}{$\ldots\left.\vphantom{\rule{0mm}{40\unitlength}}\right\rbrace^*$}
    \\[1em]
    \raisebox{30\unitlength}{\text{\footnotesize(b')}\quad
      $\left\lbrace\vphantom{\rule{0mm}{40\unitlength}}\right.$}
    {\scriptsize
  \raisebox{-15\unitlength}[70\unitlength][0mm]{\input{feyndiags/VSW}}
}
    \raisebox{30\unitlength}{$\ldots\left.\vphantom{\rule{0mm}{40\unitlength}}\right\rbrace\times$}
    &\raisebox{30\unitlength}{$\left\lbrace\vphantom{\rule{0mm}{40\unitlength}}\right.$}
    {\scriptsize
  \raisebox{-15\unitlength}[70\unitlength][0mm]{\input{feyndiags/BS}}
}
    \raisebox{30\unitlength}{$\ldots\left.\vphantom{\rule{0mm}{40\unitlength}}\right\rbrace^*$}
    \\[1em]
    \raisebox{30\unitlength}{\text{\footnotesize(c')}\quad
      $\left\lbrace\vphantom{\rule{0mm}{40\unitlength}}\right.$}
    {\scriptsize
  \raisebox{-15\unitlength}[70\unitlength][0mm]{\input{feyndiags/VSS}}
}
    \raisebox{30\unitlength}{$\ldots\left.\vphantom{\rule{0mm}{40\unitlength}}\right\rbrace\times$}
    &\raisebox{30\unitlength}{$\left\lbrace\vphantom{\rule{0mm}{40\unitlength}}\right.$}
    {\scriptsize
  \raisebox{-15\unitlength}[70\unitlength][0mm]{\input{feyndiags/BW}}
}
    \raisebox{30\unitlength}{$\ldots\left.\vphantom{\rule{0mm}{40\unitlength}}\right\rbrace^*$}
  \end{align*}}
  \caption{The virtual corrections of $\order{\alphas^2\alphaw}$
    illustrated in terms of interference diagrams of generic Feynman
    graphs (a--c) and a set of sample diagrams below (a'--c'). 
    The white circles and the double-circles in the interference diagrams
    represent tree-level and one-loop subgraphs, respectively.}
  \label{fig:virt}
\end{figure}

\begin{figure}
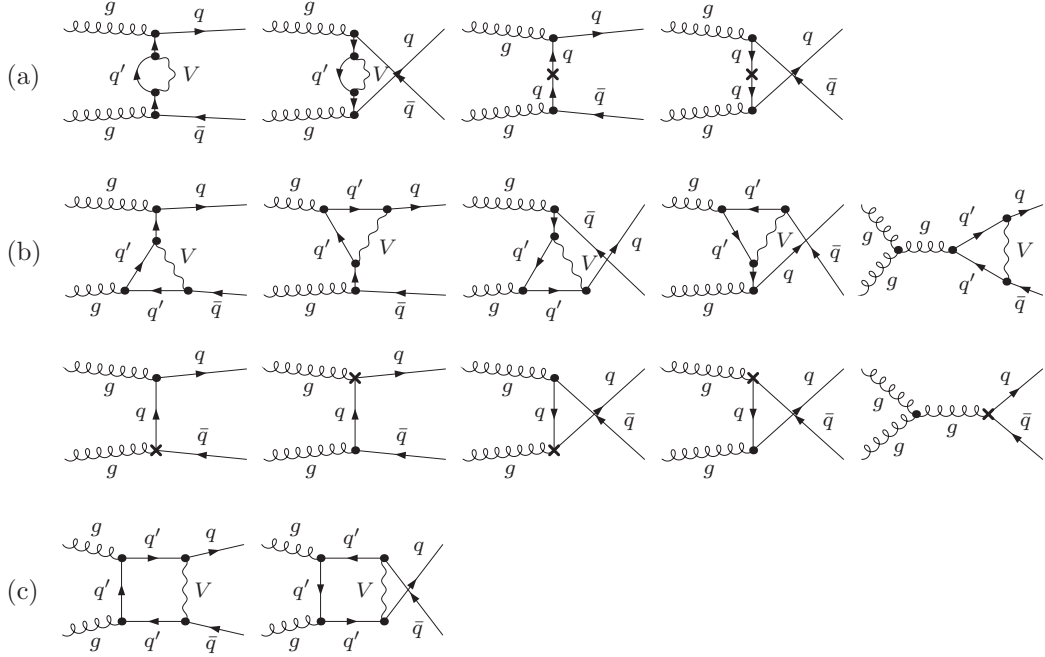

  \centering
  \unitlength=0.75bp
  \scriptsize
  \begin{minipage}{530\unitlength}
    \raisebox{30\unitlength}{\footnotesize(a)}~
  \raisebox{-15\unitlength}[70\unitlength][0mm]{\input{feynarts/V2G2Qself}}
\\[1.5em]
    \raisebox{30\unitlength}{\footnotesize(b)}~
  \raisebox{-15\unitlength}[70\unitlength][0mm]{\input{feynarts/V2G2Qvert_1}}
\\[1em]
    \raisebox{30\unitlength}{\footnotesize\phantom{(b)}}~
  \raisebox{-15\unitlength}[70\unitlength][0mm]{\input{feynarts/V2G2Qvert_2}}
\\[1.5em]
    \raisebox{30\unitlength}{\footnotesize(c)}~
  \raisebox{-15\unitlength}[70\unitlength][0mm]{\input{feynarts/V2G2Qbox}}

  \end{minipage}
  \caption{
    One-loop and counterterm diagrams for the process class
    \eqref{eq:2g2q} grouped into the self-energy (a), vertex (b), and
    box (c) corrections. All counterterms
    are restricted to the order $\order{\alphaw}$.
    $V$ denotes the vector bosons $\PW$ and $\PZ$,
    and $q^\prime$ the weak-isospin partner of $q$ for $V=\PW$ and
    $q^\prime =q$ for $V=\PZ$.}
  \label{fig:virt2g2q}
\end{figure}

\begin{figure}
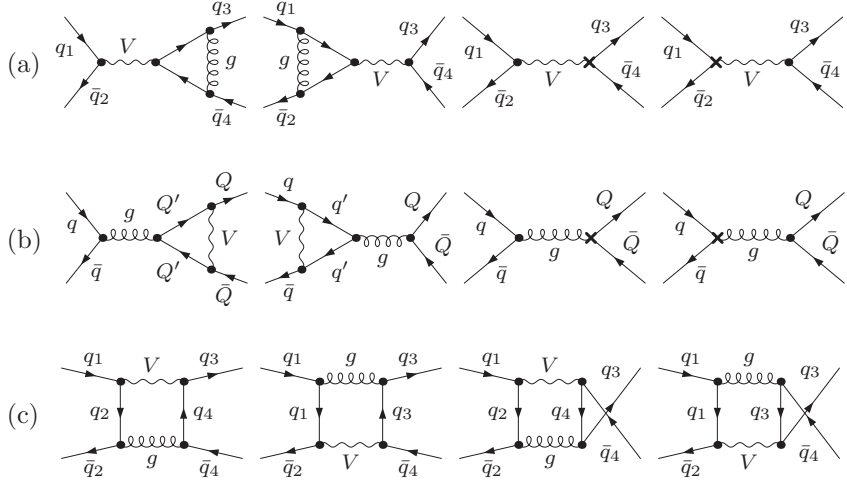

  \centering
  \unitlength=0.75bp
  \scriptsize
  \begin{minipage}{430\unitlength}
    \raisebox{30\unitlength}{\footnotesize(a)}~
  \raisebox{-15\unitlength}[70\unitlength][0mm]{\input{feynarts/VSW4Qvert1}}
\\[1.5em]
    \raisebox{30\unitlength}{\footnotesize(b)}~
  \raisebox{-15\unitlength}[70\unitlength][0mm]{\input{feynarts/VSW4Qvert2}}
\\[1.5em]
    \raisebox{30\unitlength}{\footnotesize(c)}~
  \raisebox{-15\unitlength}[70\unitlength][0mm]{\input{feynarts/VSW4Qbox}}

  \end{minipage}
  \caption{
    One-loop and counterterm diagrams for the processes
    \eqref{eq:subproc}(d--f) of $\order{\alphas\alphaw}$ grouped into
    vertex (a,b) and box (c) corrections. The triangle insertions
    are further subdivided into QCD (a) and weak (b) corrections,
    and consequently the associated counterterms are restricted to the
    order $\order{\alphas}$ and $\order{\alphaw}$, respectively.
    $V$ denotes the vector bosons $\PW$ and $\PZ$.}
  \label{fig:virt4qws}
\end{figure}

\begin{figure}
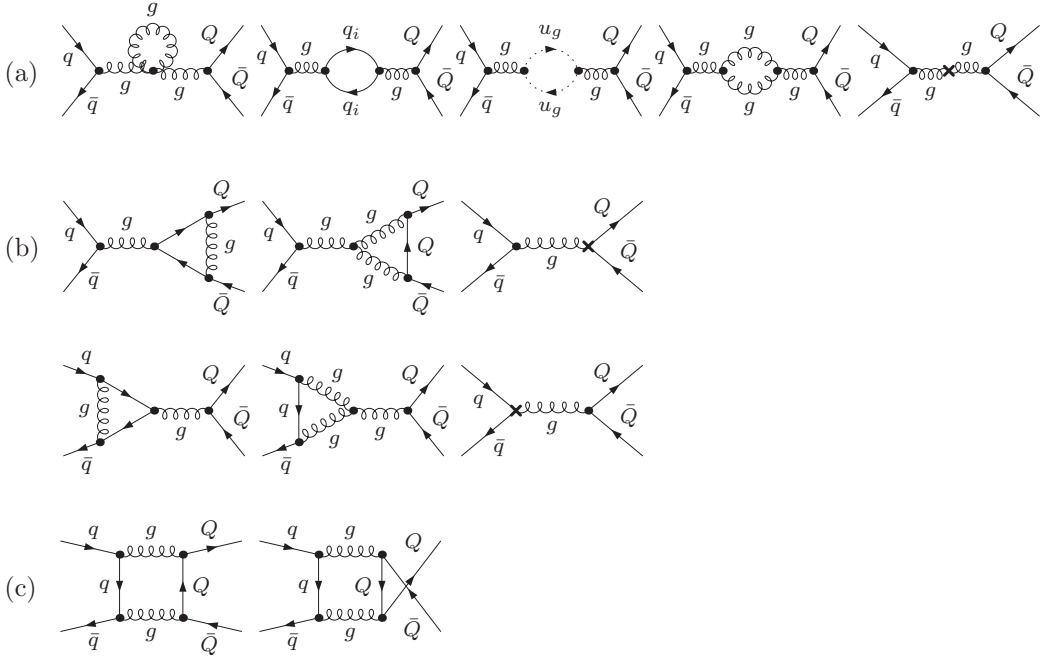

  \centering
  \unitlength=0.75bp
  \scriptsize
  \begin{minipage}{530\unitlength}
    \raisebox{30\unitlength}{\footnotesize(a)}~
  \raisebox{-15\unitlength}[70\unitlength][0mm]{\input{feynarts/VSS4Qself}}
\\[1.5em]
    \raisebox{30\unitlength}{\footnotesize(b)}~
  \raisebox{-15\unitlength}[70\unitlength][0mm]{\input{feynarts/VSS4Qvert_1}}
\\[1em]
    \raisebox{30\unitlength}{\footnotesize\phantom{(b)}}~
  \raisebox{-15\unitlength}[70\unitlength][0mm]{\input{feynarts/VSS4Qvert_2}}
\\[1.5em]
    \raisebox{30\unitlength}{\footnotesize(c)}~
  \raisebox{-15\unitlength}[70\unitlength][0mm]{\input{feynarts/VSS4Qbox}}

  \end{minipage}
  \caption{One-loop and counterterm diagrams for the processes
    \eqref{eq:subproc}(d--f) of $\order{\alphas^2}$ grouped into
    self-energy (a), vertex (b), and box (c) corrections.
    All counterterms are restricted to the order $\order{\alphas}$.}
  \label{fig:virt4qss}
\end{figure}

The virtual corrections consist of the one-loop diagrams and the
corresponding counterterms.
Because we are restricting our NLO calculation to the order
$\order{\alphas^2\alphaw}$, only the interference terms shown in
Fig.~\ref{fig:virt} are relevant.
The generic diagram for the virtual corrections to the process class
\eqref{eq:2g2q} is shown in Fig.~\ref{fig:virt}(a).
The corrections constitute the purely weak
$\order{\alphaw}$ correction to the LO $\order{\alphas^2}$ cross
section; some representative diagrams are depicted in
Fig.~\ref{fig:virt}(a').
In case of the process classes (\ref{eq:subproc}d--f) with four
external quarks, however, we have LO amplitudes of the order
${\alphaw}$ and ${\alphas}$. 
This leads to the two types of virtual corrections shown in
Figs.~\ref{fig:virt}(b,c). 
Here, we can identify genuine QCD corrections, such as the
contributions shown in Fig.~\ref{fig:virt}(c') and the first vertex
correction in Fig.~\ref{fig:virt}(b'), and weak corrections, such as the
second vertex correction in Fig.~\ref{fig:virt}(b').
However, there are also contributions such as the box diagram in
Fig.~\ref{fig:virt}(b') that cannot be assigned uniquely to
QCD nor to weak corrections.
This indicates that the separation of the ``QCD corrections'' from the
``weak corrections'' is not properly defined and instead, one must
treat them together as a whole, defined by the order in perturbation
theory. 
The contributions we referred to by the ``QCD corrections'' above
contain infrared divergences and, therefore, the real emission of an
additional gluon must be considered, which will be discussed in the
next section.
A more complete set of one-loop and counterterm diagrams
associated with the schematic illustrations in
Figs.~\ref{fig:virt}(a),~\ref{fig:virt}(b), and~\ref{fig:virt}(c) can
be found in 
Figs.~\ref{fig:virt2g2q},~\ref{fig:virt4qws}, and~\ref{fig:virt4qss},
respectively.%
\footnote{In the case with external bottom quarks there
  exist additional diagrams in Fig.~\ref{fig:virt2g2q} and
  Fig.~\ref{fig:virt4qws}(b) due to the non-vanishing mass of the
  weak-isospin partner $\Pt$, where in place of $\PW^\pm$ a charged
  would-be Goldstone boson $\phi^\pm$ is exchanged.}

Ultraviolet (UV) divergences are regularized dimensionally.
For the IR singularities also dimensional regularization is
used, but our second calculation optionally employs infinitesimal
masses as regulators.
It has been shown in the appendix of Ref.~\cite{Bredenstein2008a}
that rational terms of IR origin cancel in any unrenormalized scattering
amplitude, so that they need not be further considered in the calculation
of the one-loop amplitudes. 
The only remaining source of rational terms of IR origin involve the
wave-function renormalization constants, which were calculated
separately and are given below.

The external fields are renormalized in the on-shell scheme and hence,
all self-energy corrections to the external (on-shell) legs vanish and
can be omitted already at the level of diagram generation. 
The renormalization of the strong coupling constant is done using the
$\MSbar$ scheme for the $N_f=5$ light quarks and by subtracting
the contribution of the heavy top-quark loop in the gluon self-energy
at zero momentum transfer. 
Therefore, the running of the strong coupling constant is driven by
the five light quark flavours only.

Similar to the loop diagrams, the renormalization constants need to be
evaluated at different orders, where the ones of
$\order{\alphas}$ are 
\begin{align}
  \label{eq:dZgs}
  \delta Z_{\gs} \rvert_{\order{\alphas}}
  &= -\frac{\alphas}{4\pi}\left[
    \left(\frac{11}{2}-\frac{N_f}{3}\right)
    \left(\Deltauv+\ln\left(\frac{\mu^2}{\mu_\rR^2}\right)\right)
    -\frac{1}{3}B_0(0,m_\Pt,m_\Pt)
  \right],\\
  \delta Z_{G} \rvert_\order{\alphas}
  &= \frac{\alphas}{2\pi}\left[
    \left(\frac{5}{2}-\frac{N_f}{3}\right)B_0(0,0,0)
    -\frac{1}{3}B_0(0,m_\Pt,m_\Pt)
  \right],\\
  \delta Z_{q} \rvert_\order{\alphas}
  &= -\frac{\alphas}{3\pi}B_0(0,0,0),
\end{align}
and the ones of $\order{\alphaw}$ are
\begin{align}
  \delta Z_{q}^\rL \rvert_{\order{\alphaw}}
  &= \frac{\alpha}{4\pi}\left[
    (g_q^-)^2 \left(1+2B_1(0,0,\mu_\PZ)\right)
    +\frac{1}{2\sw^2} \left(1+2B_1(0,0,\mu_\PW)\right)
  \right],\\
  \delta Z_{q}^\rR \rvert_{\order{\alphaw}}
  &= \frac{\alpha}{4\pi}
    (g_q^+)^2 \left(1+2B_1(0,0,\mu_\PZ)\right),
  \label{eq:dZq}
\end{align}
where our notation for the 2-point functions $B_{0,1}$ follows
Ref.~\cite{Denner2006}, $\mu$ is the arbitrary reference mass
of dimensional regularization, $\mu_\rR$ the renormalization scale, and
\begin{equation}
\Deltauv = \frac{2}{4-D} - \gamma_{\mathrm{E}} + \ln(4\pi)
\end{equation}
denotes the standard one-loop UV divergence in $D$ dimensions.
Here $\delta Z_{q}^{\rR/\rL}$ and $\delta Z_{G}$ are the field-renormalization
constants of the right/left-handed quark fields and of the gluon field,
respectively, and $\delta Z_{\gs}$ connects the bare
($\gso$) and the renormalized ($\gs$) strong coupling constant,
$\gso=(1+\delta Z_{\gs})\gs$.
The couplings $g_q^\pm$ are defined via the 3rd component of weak-isospin,
$I^3_{\mathrm{w},q}$, and the electric charge $Q_q$ of the quark $q$,
\begin{equation}
  \label{eq:9}
  g_q^- = \frac{I^3_{\mathrm{w},q}-\sw^2 Q_q}{\sw\cw},\quad 
  g_q^+ = -\frac{\sw}{\cw}Q_q,\quad
  (I^3_{\mathrm{w},q},Q_q) =
  \begin{cases}
    (+1/2,+2/3), \quad q=u_i, \\
    (-1/2,-1/3), \quad q=d_i.
  \end{cases} 
\end{equation}

At this point some comments on the use of complex masses and couplings,
as dictated by the complex-mass scheme~\cite{Denner1999,Denner2005a},
are appropriate. 
This scheme, which was primarily introduced to achieve a consistent, gauge-invariant
description of gauge-boson resonances at LO and NLO, does not only comprise
the consistent use of complex parameters in amplitudes, but
also complex generalizations of the renormalization constants for
the complex masses and couplings as compared to on-shell
renormalization schemes for real masses (see \eg Ref.~\cite{Denner1993}).
Note, however, that the order $\order{\alphas^2\alpha}$ of the corrections
considered in this calculation does not involve weak corrections to gauge-boson
propagators and weak couplings, so that the complex generalization of the
relevant renormalization constants given in Eqs.~(\ref{eq:dZgs})--(\ref{eq:dZq})
just concerns the insertion of complex masses and weak couplings.

\subsubsection{Real corrections}
\label{sec:real}

\begin{figure}
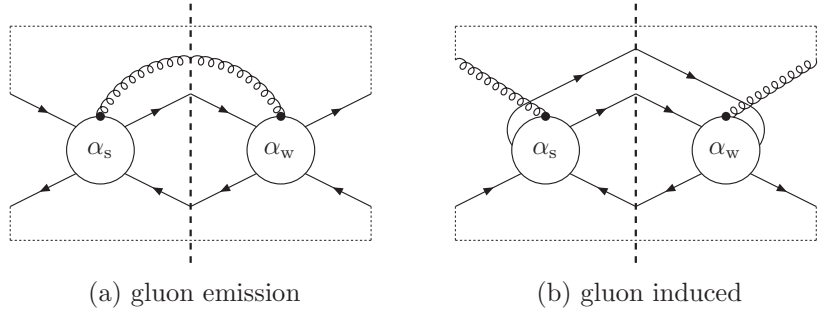

  \centering
  \begin{tabular}{cc}
    \scalebox{\myscale}{\input{feyndiags/IntR}} &
    \scalebox{\myscale}{\input{feyndiags/IntRg}} \\
    {\footnotesize(a) gluon emission} & 
    {\footnotesize(b) gluon induced}
  \end{tabular}
  \caption{
    The real corrections at $\order{\alphas^2\alphaw}$
    illustrated in terms of interference diagrams of generic Feynman
    graphs. The circles represent tree-level subgraphs.}
  \label{fig:cutreal}
\end{figure}

The real corrections receive contributions from the partonic processes
that are illustrated in terms of interference diagrams in
Fig.~\ref{fig:cutreal}. 
The contribution shown in Fig.~\ref{fig:cutreal}(a) corresponds to the
additional emission of a gluon from the processes
(\ref{eq:subproc}d--f), properly taking into account only the
interference terms that contribute at the order
$\order{\alphas^2\alphaw}$. 
The gluon-induced corrections in Fig.~\ref{fig:cutreal}(b) are obtained from
the preceding by crossing the gluon into the initial state.

The real-emission cross section contains infrared divergences
in the phase-space integration which have their origin in the regions 
where a final-state parton becomes soft or collinear to another
parton. 
The soft and the final-state collinear singularities cancel against
the corresponding singularities in the virtual corrections for
sufficiently inclusive observables by virtue of the
Kinoshita--Lee--Nauenberg theorem \cite{Kinoshita1962,Lee1964}.
The remaining initial-state collinear singularities are process
independent and absorbed into the NLO PDFs by QCD factorization, 
which is technically accomplished by subtracting a so-called collinear
counterterm ($\rd\sigC$) from the NLO cross section.
The subtraction formalism reshuffles the IR singularities by
constructing a subtraction term ($\rd\sigA$) to the real correction
($\rd\sigR$) which mimics its singular behaviour to render their
difference integrable in four dimensions.  
The subtraction term can be integrated analytically in $D=4-2\epsilon$
dimensions over the singular one-particle subspace, generating
$1/\epsilon$ and $1/\epsilon^2$ poles that cancel against the
corresponding poles in the virtual corrections ($\rd\sigV$) and the
collinear counterterm.

To this end, we employ the Catani--Seymour dipole subtraction 
formalism~\cite{Catani1997}, which constructs the subtraction term
in terms of so-called dipoles, which are built from the LO amplitudes
($\sigB$) and dipole operators ($\CSV$) which in general contain
colour and helicity correlations. 
The NLO contribution to the hard-scattering cross section from
Eq.~\eqref{eq:11} can be schematically written as, 
\begin{align}
  \label{eq:12}
  \sigloop
  &= \int_3\rd\sigR + \int_2\rd\sigV + \int_2\rd\sigC \nnnl
  &= \int_3\;\biggl[ \bigl(\rd\sigR\bigr)_{\epsilon=0} 
  - \bigl(\rd\sigA\bigr)_{\epsilon=0} \biggr]
  + \int_2\;\biggl[ \rd\sigV + \rd\sigC
  + \int_1 \rd\sigA\biggr]_{\epsilon=0}. \nnnl
  &= \int_3\;\biggl[ \bigl(\rd\sigR\bigr)_{\epsilon=0} 
  - \biggl(\sum_\text{dipoles}\;\rd\sigB\otimes\CSV\biggr)_{\epsilon=0}
  \biggr] \nnnl
  &\quad + \int_2\;\biggl[ \rd\sigV 
  + \rd\sigB\otimes\CSI \biggr]_{\epsilon=0}  
  + \int_0^1\rd x\int_2\rd\sigB\otimes\left(\CSK+\CSP\right),
\end{align}
where $\int_m$ denotes the integration over the $m$-particle phase
space, $\otimes$ encodes the possible colour and helicity
correlations between $\rd\sigB$ and the dipole operators, and the
integration over $x$ corresponds to a convolution over the momentum
fraction of the incoming partons.
The insertion operators $\CSI$, $\CSK$, and $\CSP$ emerge from the
collinear counterterm and the integration of the subtraction term
over the singular one-particle subspace. 
More details and the explicit expressions for $\CSI$, $\CSK$, and $\CSP$
can be found in Ref.~\cite{Catani1997}.

\begin{figure}
  \centering
  \scalebox{\myscale}{\input{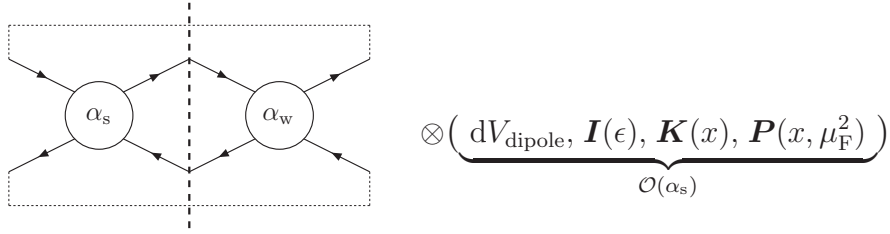}}
  \raisebox{11.5mm}{
    $\otimes\bigl(\underbrace{\text{
        $\CSV$,\;$\CSI$,\;$\CSK$,\;$\CSP$
      }}_{\order{\alphas}}\bigr)$ 
  }
  \caption{
    The building blocks for the
    subtraction terms at $\order{\alphas^2\alphaw}$
    illustrated in terms of interference diagrams of generic Feynman
    graphs. The circles represent tree-level subgraphs.
    The symbol $\otimes$ denotes possible colour and helicity
    correlations between the dipole operators and the LO
    amplitudes.}
  \label{fig:cutdipole}
\end{figure}
The dipole operators introduced above are all of $\order{\alphas}$,
which in turn forces us to restrict the calculation of the
colour-correlated LO amplitudes to the interference terms of
$\order{\alphas\alphaw}$.
This is schematically illustrated in Fig.~\ref{fig:cutdipole}. 

For the second calculation which used both dimensional and mass
regularization for the IR divergences, 
the generalization of the above formalism for
massive partons~\cite{Catani2002} is used for the latter scheme.

\section{Numerical results}
\label{sec:num}

\subsection{Input parameters and setup}
\label{sec:input}

For the numerical evaluation we use the input parameters of
Ref.~\cite{Amsler2008},
\begin{align}
  G_\mu &= 1.16637\times 10^{-5}~\GeV^{-2}, 
  & \alphas(M_\PZ) 
  &= 0.129783, \nnnl*
  M_\PW^\mathrm{OS} &= 80.398~\GeV, 
  & M_\PZ^\mathrm{OS} &= 91.1876~\GeV,\nnnl*
  \Gamma_\PW^\mathrm{OS} &= 2.141~\GeV, 
  & \Gamma_\PZ^\mathrm{OS} &= 2.4952~\GeV, \nnnl*
  m_\Pt &= 172.5~\GeV. & 
%M_\PH &= 125~\GeV.
  \label{eq:4}
\end{align}
Note that there is no Higgs-boson mass dependence in the predictions for
dijet production at the considered order.
The complex-mass scheme for the $\PW$ and $\PZ$ bosons corresponds
to a fixed-width description of the resonance and requires a
conversion of the on-shell gauge-boson masses to the pole masses $M_V^\mathrm{OS}$
as follows~\cite{Bardin1988},
\begin{equation}
  \label{eq:7}
  M_V = \frac{M_V^\mathrm{OS}}{c_V},\quad 
  \Gamma_V = \frac{\Gamma_V^\mathrm{OS}}{c_V}, \quad
  c_V = \sqrt{1+\left(\frac{\Gamma_V^\mathrm{OS}}{M_V^\mathrm{OS}}\right)^2},
\end{equation}
giving
\begin{align}
  M_\PW &= 80.3695\ldots\GeV, 
  & M_\PZ &= 91.1535\ldots\GeV,\nnnl*
  \Gamma_\PW &= 2.1402\ldots\GeV, 
  & \Gamma_\PZ &= 2.4943\ldots\GeV.
\end{align}

For the PDFs we use the CTEQ6L1~\cite{Pumplin2002a} set,
which dictates the value of $\alphas(M_\PZ)$ in Eq.~\eqref{eq:4}.  
A consistent QCD calculation, of course, requires the use
of LO, NLO, or NNLO PDFs for the calculation of the respective cross-section
prediction. Our aim here is, however, to provide a relative correction
factor for weak effects that is to be applied to state-of-the-art QCD 
predictions. This factorization procedure is better motivated than just adding
QCD and electroweak corrections because of
the factorization of IR-sensitive corrections. 
The weak correction factor has to be derived from a single PDF set, but 
is rather insensitive to PDFs.%
\footnote{The consistent use of LO PDFs is fully justified for 
the electroweak tree-level contributions and
the weak loop corrections (containing the large weak Sudakov logarithms) to the QCD channels. 
The remaining part of the calculated correction actually has the character of
a QCD correction, viz.\ the interference of QCD loops with
weak LO diagrams and the corresponding real corrections, strictly requiring NLO PDFs. 
Recall, however, that the two different
loop contributions cannot be separated, as explained in Sect.~\ref{sec:nlo}.
Since the calculated corrections are generally rather moderate,
our procedure is certainly acceptable within the remaining uncertainty 
of the aimed combination of QCD and weak corrections.} 
The renormalization scale is set equal to the factorization scale,
chosen as the transverse momentum of the leading jet
\begin{equation}
  \label{eq:8}
  \mu_\rR = \mu_\rF \equiv \mu = \ktl.
\end{equation}

\subsection{Phase-space cuts and event selection}
\label{sec:cuts}

The definition of an IR-safe jet observable requires the
recombination of soft and/or collinear partons in the final state, as
well as constraining the phase space by imposing cuts. 
The jets emerge from the final-state partons via the anti-$k_\rT$
algorithm~\cite{antikt}, where we have set the angular separation
parameter to $R=0.6$. 
Recombination is performed using four-momentum summation. 

We require the jets to have transverse momenta $\kti$
larger than $\ktcut$ and demand them to be central
by restricting their rapidities $y_i$ to the range $\lvert y_i\rvert < \ycut$ with
the values
\begin{equation}
  \label{eq:5}
   \ycut= 2.5, \qquad
   \ktcut = 25~\GeV.
\end{equation}

\subsection{Results}
\label{sec:res}

In the following, we present the numerical results for dijet production
at the LHC, \ie for a $\Pp\Pp$ initial state, at the centre-of-mass (CM)
energies of $\sqrt{s}=7~\TeV$, $8~\TeV$, and $14~\TeV$,
and at the Tevatron,  \ie for $\Pp\bar\Pp$ collisions at the CM
energy of $\sqrt{s}=1.96~\TeV$.

We denote the full LO cross section
through $\order{\alphas^2,\,\alphas\alpha,\,\alpha^2}$
by $\sigtree$ and define the NLO corrections relative to the LO cross
section via
\begin{equation}
  \label{eq:25}
  \sigloop = \sigtree \times (1+\delloop).
\end{equation}
Furthermore, we denote the LO QCD cross section of
$\order{\alphas^2}$ by $\sigqcd$ and introduce a correction factor
for the left-over LO contributions of order $\alphas\alpha$ and
$\alpha^2$,
\begin{equation}
  \label{eq:26}
  \sigtree = \sigqcd\times(1+\deltree).
\end{equation}
For the NLO cross section this leads to
\begin{align}
  \label{eq:27}
  \sigloop &= \sigqcd\times(1+\deltree)\times (1+\delloop) \nnnl
  &\simeq\sigqcd\times(1+\deltree+\delloop).
\end{align}
With respect to the LO QCD cross section, the total correction is
given by the sum $\delloop+\deltree$.
Owing to the rather moderate size of the corrections,
the difference in defining the NLO corrections $\delloop$
relative to $\sigtree$ or $\sigqcd$ constitutes a higher-order effect
which is negligible.

\subsubsection{The dijet invariant mass at the LHC}
\label{sec:mjj}

\begin{table}
  \centering
  \footnotesize
  \input{tables/cdj7totmjj.tex}\\
  \input{tables/cdj8totmjj.tex}\\
  \input{tables/cdj14totmjj.tex}
  \caption{Integrated dijet cross sections and respective corrections
    for various ranges of the
    dijet invariant mass $\mjj$ at the LHC with CM energies 7~\TeV,
    8~\TeV, and 14~\TeV.}
  \label{tab:mjjcuts}
\end{table}
The dijet invariant mass is defined as 
$\mjj=\sqrt{(E_1+E_2)^2-(\mathbf{p}_1+\mathbf{p}_2)^2}$,
where $E_{1,2}$ and $\mathbf{p}_{1,2}$ denote the energies and the
momenta of the leading and subleading jets, respectively.
In Table~\ref{tab:mjjcuts} we present the integrated LO cross sections
$\sigtree$, $\sigqcd$, the NLO cross section $\sigloop$, and the
relative correction factors $\delloop$, $\deltree$,
$\delsumabbr=\delsum$
for different cuts on the dijet invariant mass $\mjj$ at the LHC. 
We note that the lowest cut value of $\mjj>50~\GeV$ is already covered by
the standard setup via the cut on the transverse momenta of the jets,
$\ktcut = 25~\GeV$, and does not constitute a further
restriction on the cross section. 

The cross section is dominated by the region close to
the cut, which is reflected by the rapid decrease of the integrated
cross section with increasing values for the cut on $\mjj$.
Therefore, also the corrections are dominated by the region given by
the lowest accepted $\mjj$ values.

Comparing the two cross sections $\sigtree$ and $\sigqcd$, we find
that the LO cross section with minimal cuts
is predominantly given by the QCD cross
section and that the electroweak effects ($\deltree$) typically stay
below the per-cent level.
However, we observe a steady increase in $\deltree$ with higher $\mjj$,
which can be explained by the parton luminosities at the LHC:
\begin{figure}
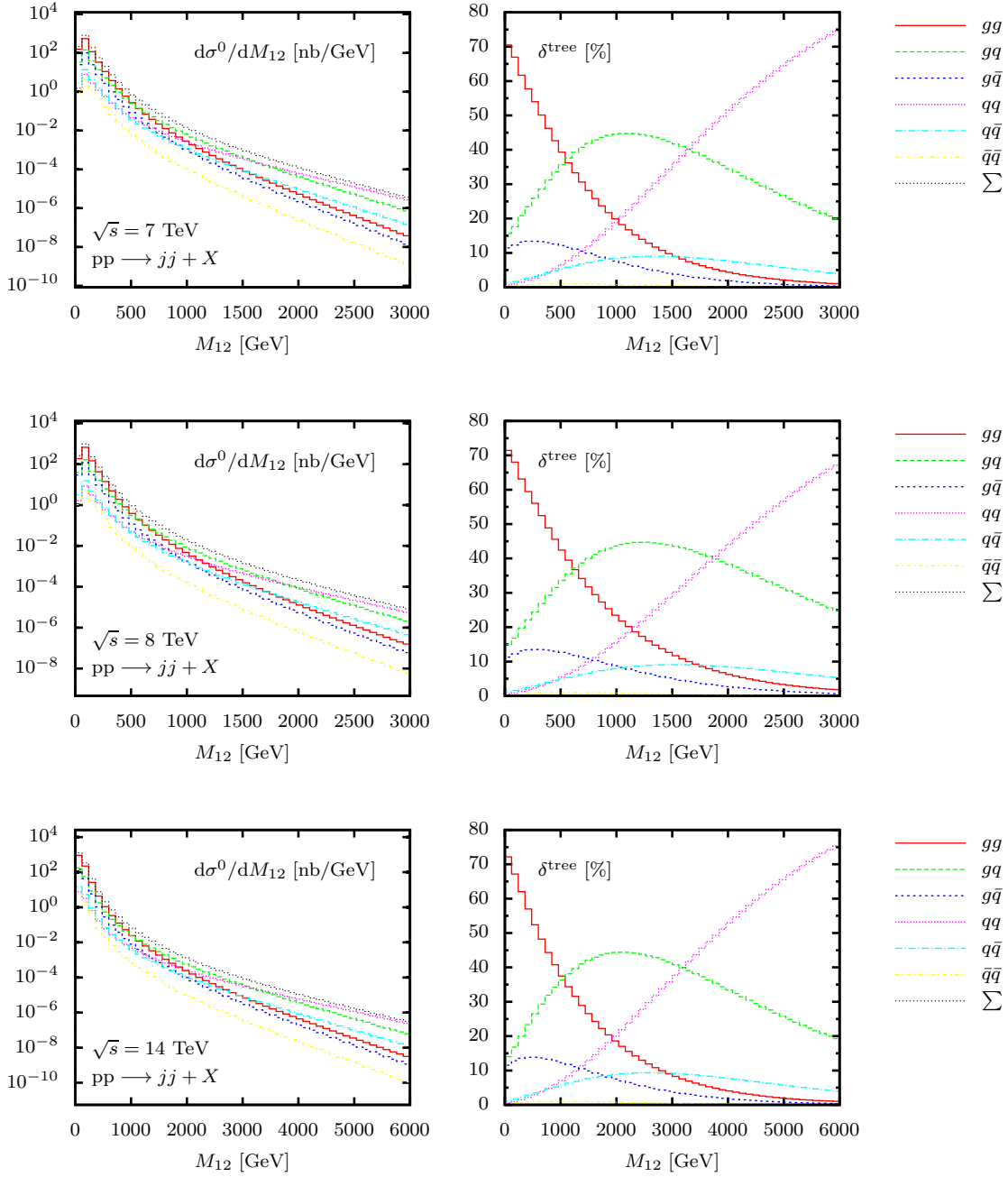

  \centering
  \scriptsize
  \input{plots/cdj7dist_MJJ_pids.tex}\\
  \input{plots/cdj8dist_MJJ_pids.tex}\\
  \input{plots/cdj14dist_MJJ_pids.tex}
  \caption{The LO contributions to the dijet invariant mass $\mjj$
    distribution from the different initial-state parton combinations
    at the LHC with CM energies 7~\TeV, 8~\TeV, and 14~\TeV. 
    Left: absolute predictions; right: relative contributions
    $\delta^\text{tree}$.}
  \label{fig:mjjpids}
\end{figure}
As can be seen from Fig.~\ref{fig:mjjpids}, the LO cross section is
dominated at lower $\mjj$ by the $gg$- and $gq$-initiated processes,
for which $\deltree$ vanishes. 
At higher $\mjj$, the $qq$-initiated processes with $\deltree\ne0$
become dominant, leading to the behaviour described above. 
The running of the strong coupling also acts in favour of increasing
$\deltree$ with higher cuts on $\mjj$.
Comparing the distributions in Fig.~\ref{fig:mjjpids} for
$\sqrt{s}=7~\TeV$ and $14~\TeV$ we observe a shift of
the transition region from $gg$- to $qq$-domination from
$\sim1.5~\TeV$ to $\sim3~\TeV$, respectively.  
This trend is due to the fact that lower values of $\sqrt{s}$ require
larger partonic momentum fractions $x$ for a fixed value of $\mjj$, and
the (valence) quark PDFs are enhanced over the gluon PDF for larger $x$.
This also explains the $\sqrt{s}$ dependence of the EW contribution
$\deltree$, which decreases
with higher CM energies for the same cut on $\mjj$.

The purely weak corrections are negative throughout and increase in
magnitude from $-0.02\%$ to $-3.6\%$ in case of the $\sqrt{s}=7~\TeV$
setup for a $\mjj$ cut of $50~\GeV$ and $3~\TeV$, respectively.
This behaviour partly originates from the corrections containing 
weak logarithms $\ln\left(\frac{M_\PW^2}{Q^2}\right)$,
which become larger by effectively restricting the cross
section to higher energy scales $Q$ via increasing cuts
on the invariant mass.

Compared to $\deltree$, the relative weak corrections $\delloop$
show a far weaker dependence on the CM energy.
The corrections $\deltree$ and $\delloop$ are similar in magnitude,
but of opposite sign, leading to large cancellations in the sum.

Note that the weak loop corrections $\delloop$ in the $\TeV$ range are
small in comparison to the typical size of the EW Sudakov factor 
$\frac{\alphaw}{\pi}\ln^2\left(\frac{M_\PW^2}{Q^2}\right)$, 
which amounts to tens of per cent. 
This is due to the fact that also for large invariant mass
$\mjj$ the cross section is not dominated by the Sudakov regime which
requires that the absolute values of both partonic Mandelstam variables 
$\shat=(p_a+p_b)^2$ and $\that=(p_a-k_c)^2$ 
are much larger than $M_\PW^2$.
Instead, the cross section here is dominated by the Regge (forward)
region where $\shat$ is large, but $|\that|$ remains small.

\begin{figure}
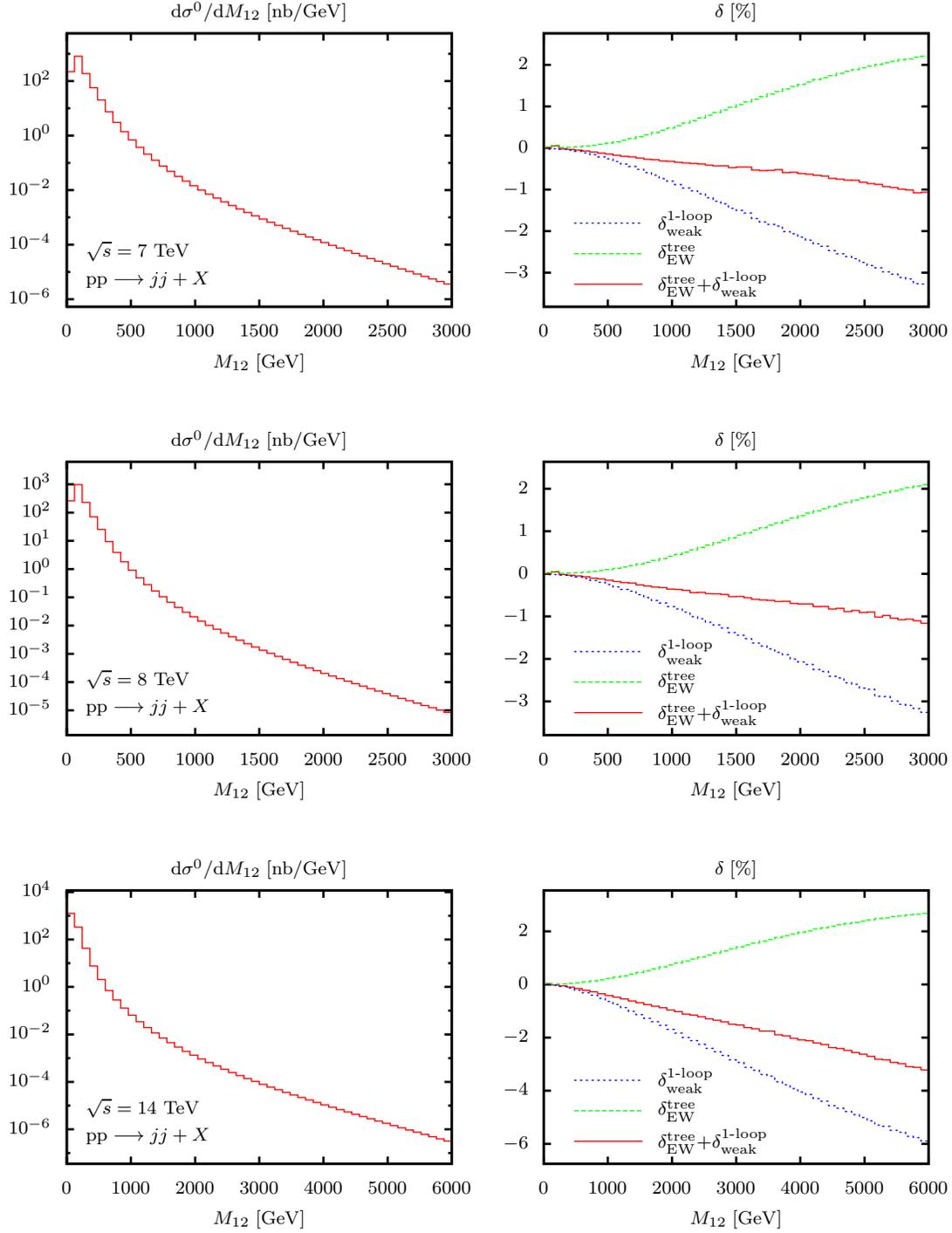

  \centering
  \scriptsize
  \input{plots/cdj7dist_MJJ.tex}\\
  \input{plots/cdj8dist_MJJ.tex}\\
  \input{plots/cdj14dist_MJJ.tex}  
  \caption{Differential distributions with respect to the 
    dijet invariant mass $\mjj$ at the LHC with CM energies 7~\TeV,
    8~\TeV, and 14~\TeV.
    Left: absolute predictions; right: relative contributions $\delta$.}
  \label{fig:delmjj}
\end{figure}
This feature is also evident in the dijet invariant-mass
distributions shown in Fig.~\ref{fig:delmjj}.
The relative corrections given in
Table~\ref{tab:mjjcuts} are almost identical to the corrections in
Fig.~\ref{fig:delmjj} at the respective cut value of $\mjj$, 
since the corrections are dominated
by the region close to the cut.

% \begin{figure}
%   \centering
%   \scriptsize
%   \input{plots/cdj7dist_MJJ_DBLE.tex}\\
%   \input{plots/cdj7dist_MJJ_DBLE_delta.tex}
%   \caption{Double-differential distribution with respect to the 
%     dijet invariant mass $\mjj$ and $\ystar$ at the LHC with a CM
%     energy of 7~\TeV.}
%   \label{fig:dbledelmjj7}
% \end{figure}
\begin{figure}
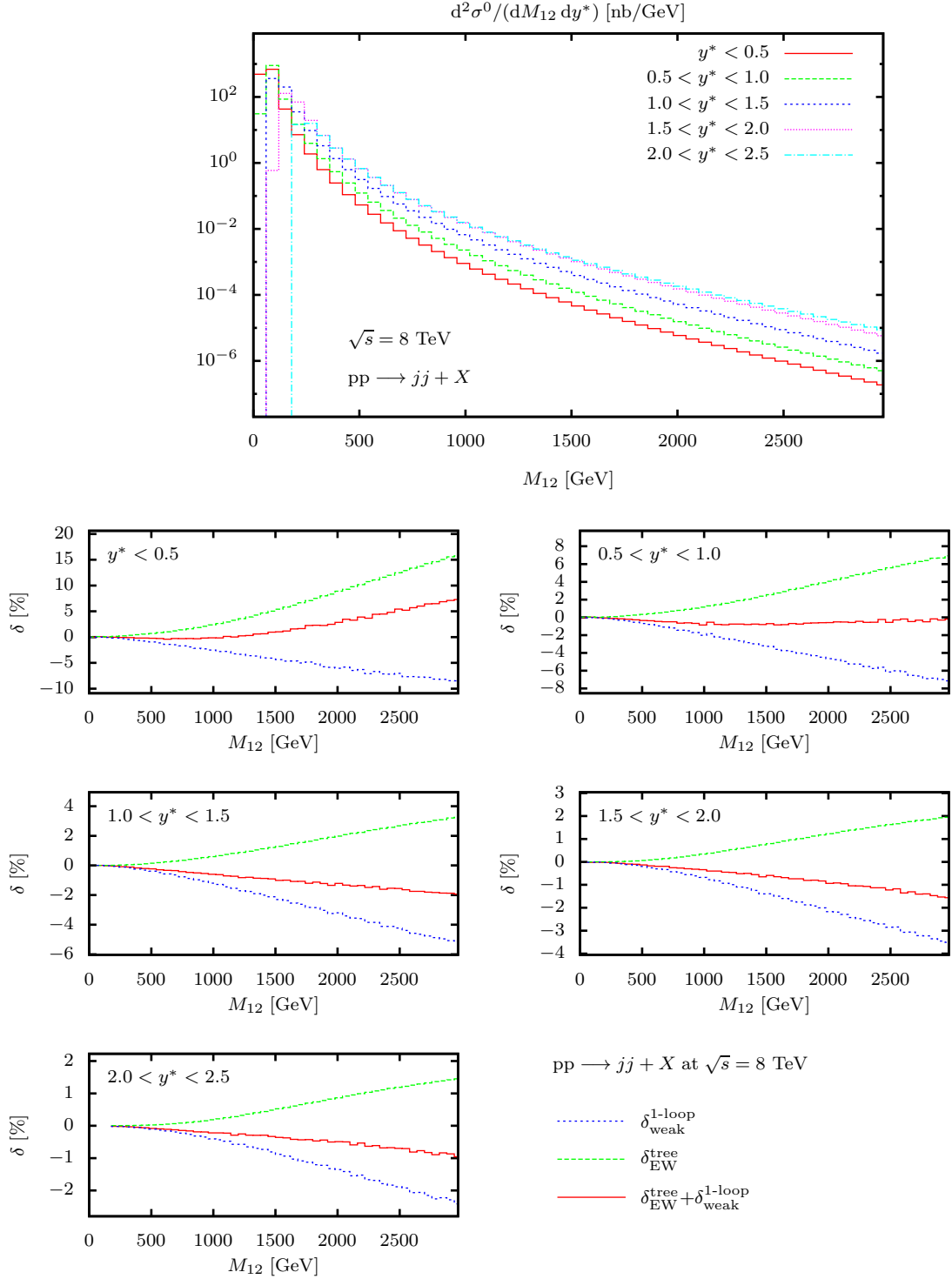

  \centering
  \scriptsize
  \input{plots/cdj8dist_MJJ_DBLE.tex}\\
  \input{plots/cdj8dist_MJJ_DBLE_delta.tex}
  \caption{Double-differential distribution with respect to the 
    dijet invariant mass $\mjj$ and $\ystar$ at the LHC with a CM
    energy of 8~\TeV. In the absolute prediction (uppermost plot)
    the cross section is divided by the bin width in $\ystar$.}
  \label{fig:dbledelmjj8}
\end{figure}
\begin{figure}
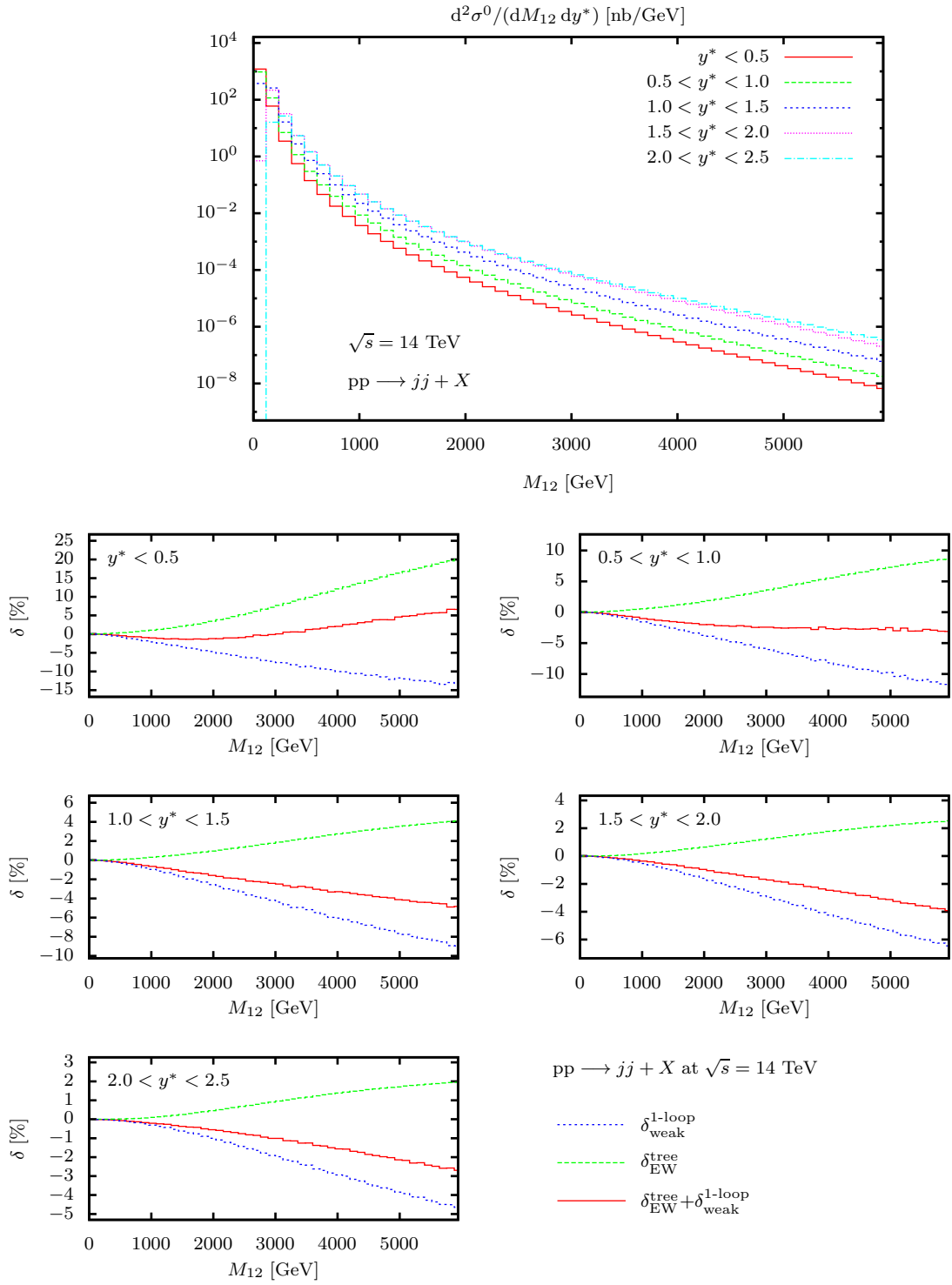

  \centering
  \scriptsize
  \input{plots/cdj14dist_MJJ_DBLE.tex}\\
  \input{plots/cdj14dist_MJJ_DBLE_delta.tex}
  \caption{Same as in Fig.~\ref{fig:dbledelmjj8}, but for a CM energy
    of 14~\TeV.}
  \label{fig:dbledelmjj14}
\end{figure}
Figures~\ref{fig:dbledelmjj8} and \ref{fig:dbledelmjj14} show
the results for the $\mjj$ distributions corresponding to different
rapidity ranges.
Specifically, the phase space is divided into different regions of
$\ystar$, which is defined as half the rapidity difference of the two
leading jets, 
\begin{equation}
  \label{eq:20}
  \ystar = \frac{\lvert y_1 - y_2 \rvert}{2}.
\end{equation}
For the binning in $\ystar$ we have chosen five bins of equal width
covering the range $0<\ystar<2.5$.
In all $\ystar$ bins we observe the above described behaviour of the
relative corrections that increase in magnitude with higher invariant
mass, where the tree-level EW contributions and the radiative weak
corrections give positive and negative contributions, respectively.
Furthermore, a strong dependence on $\ystar$ is observed, such that 
both, $\deltree$ and $\delloop$, are
larger for smaller values of $\ystar$.
For $\sqrt{s}=8~\TeV$ and the invariant mass of $3~\TeV$ the
corrections $\deltree$ and $\delloop$ in the first bin ($\ystar<0.5$)
amount to approximately $15\%$ and $-10\%$, respectively,
whereas in the highest $\ystar$-bin ($2<\ystar<2.5$) we observe about 
$1.5\%$ and $-2.5\%$ for the respective corrections.
As it is evident from the numbers quoted above, 
$\deltree$ decreases more rapidly than $\delloop$ for higher $\ystar$.
The sum of both contributions, thus,
has a positive net-contribution for $\ystar<0.5$, gradually
decreasing with higher $\ystar$ values. 
In the range of highest $\ystar$ ($2<\ystar<2.5$), even  
a negative net-contribution results.
Qualitatively we observe the same behaviour also at the
CM energy of $\sqrt{s}=14~\TeV$. 
Here we examine dijet invariant masses up to $6~\TeV$, where we
observe approximately $20\%$ and $-15\%$ in the first bin
($\ystar<0.5$), and $2\%$ and $-5\%$ in the highest $\ystar$-bin
($2<\ystar<2.5$) for $\deltree$ and $\delloop$, respectively.

As already indicated in the discussion of integrated cross sections
above, the behaviour of the corrections can be better understood by
identifying the Sudakov regime in phase space.
For the contributions with $2\rightarrow2$ kinematics, \ie the LO
cross sections and the virtual corrections, the following relations
hold: 
\begin{equation}
  \label{eq:24}
  \lvert \hat{y}_1\rvert =\lvert -\hat{y}_2\rvert = \ystar, \quad
  \hat{s} = \mjj^2, \quad
  \hat{t} = - \frac{\mjj^2}{1+\re^{\pm2\ystar}}, \quad
  \hat{u} = - \frac{\mjj^2}{1+\re^{\mp2\ystar}},
\end{equation}
where $\hat{y}_i$ and $\hat{s}$, $\hat{t}$, $\hat{u}$ denote the
rapidity and the partonic Mandelstam variables.
The two different signs refer to $\hat{y}_1\gtrless0$.
For large $\mjj\gg M_\PW$ and small $\ystar$ we reside in the
Sudakov regime, where all scales are simultaneously large compared to
the vector-boson masses. 
Here, all logarithms of the form $\alphaw\ln^2\left(Q^2/M_V^2\right)$, 
with $Q^2=\hat{s},-\hat{t},-\hat{u}$ become large, leading to the
observed enhancements of the weak radiative corrections.
As can be seen in the absolute distributions in
Figs.~\ref{fig:dbledelmjj8} and \ref{fig:dbledelmjj14}, however, this
regime delivers only a small fraction to the cross section if the
high-energy region is defined via cuts on the invariant mass $\mjj$.

Although the corrections $\deltree$ show a similar behaviour, their
origin is of a completely different nature: 
As discussed above, the bulk of the contributions to $\deltree$
originate from the $qq$-initiated processes, in particular from the
partonic subprocesses $\Pu\Pu\rightarrow\Pu\Pu$ and
$\Pu\Pd\rightarrow\Pu\Pd$. 
Owing to their colour structure,
the interference terms of $\order{\alphas\alpha}$ considered here
receive contributions from products of $t$-, $u$-, and $s$-channel diagrams
only, but not from squares of those topologies. 
Therefore, these interferences
appear to be more central than the larger LO QCD contributions which
are dominated by squared topologies.

In summary, for high $\mjj$,
where the corrections are largest, the cross section is dominated
by the contribution coming from the highest possible $\ystar$
values.
However, this region receives the smallest corrections, leading to the
small corrections that we observe in Table~\ref{tab:mjjcuts} even for
very high $\mjj$ cuts.
As we will see in the next section, this behaviour is reversed in case
of the transverse-momentum distributions.
Despite the different origin of the 
relative corrections $\delloop$ and $\deltree$, they conspire together
to large cancellations in the sum $\delsum$.
The degree of this compensation, however, depends on the chosen cuts
defining the observable.

\subsubsection{The transverse momenta of the leading and subleading jet 
at the LHC}
\label{sec:kt}

\begin{table}
  \centering
  \footnotesize
  \input{tables/cdj7totkt.tex}\\
  \input{tables/cdj8totkt.tex}\\
  \input{tables/cdj14totkt.tex}
  \caption{Integrated dijet cross sections and respective corrections
    for various ranges of the
    transverse momentum of the leading jet $\ktl$ at the LHC with CM
    energies 7~\TeV, 8~\TeV, and 14~\TeV.}
  \label{tab:ktcuts}
\end{table}
In Table~\ref{tab:ktcuts} we list the various integrated cross
sections at LO, the NLO cross section, and the correction factors
defined above for different cuts on the transverse momentum of the
leading jet, $\ktl$.  
The cut $\ktl>25~\GeV$ in the first column is already imposed by the
default set of cuts and does not represent a further restriction to
the cross section. 
Again, the integrated cross section decreases rapidly with more
restrictive cuts on the transverse momentum, indicating that the cross
section, and with it the corrections, are dominated by the region
with the lowest accepted $\ktl$.

The weak radiative corrections display the expected Sudakov-type
behaviour with increasing negative corrections for higher $\ktl$-cuts
and only a modest dependence on the CM energy of the
collider. 
For cut values of $25~\GeV$ up to $1.5~\TeV$ they increase from
$-0.02\%$ to $-6\%$ for $\sqrt{s}=7~\TeV$.
For $\sqrt{s}=14~\TeV$ and a cut of $\ktl>2.5~\TeV$ the radiative
corrections even amount to $-11\%$.
The LO EW contributions show a much stronger dependence on the
collider energy than the loop corrections, in particular 
for more restrictive cuts. 
This $\sqrt{s}$ dependence is also stronger than the one
observed for the dijet invariant-mass spectra discussed
in the previous section.

Considering that $\ktl\le\mjj/2$ at LO
and that the cross section as well as
the corrections are dominated by the region of lowest accepted $\mjj$,  
one might naively expect that the results for a fixed cut on $\ktl$
should be comparable to the respective corrections for a $\mjj$ cut of
twice that value. 
However, the region close to $\mjj\approx2\ktl$ at LO
requires central jet
production in the partonic CM frame, \ie $y^*\approx0$, so that the
cross section defined via the $\ktl$ cut is dominated by the Sudakov
regime.
This explains the smaller cross section and the larger corrections as
compared to the cross section defined via the corresponding $\mjj$ cut
discussed in the previous section.

\begin{figure}
  \centering
  \scriptsize
  \input{plots/cdj7dist_KT1.tex}\\
  \input{plots/cdj8dist_KT1.tex}\\
  \input{plots/cdj14dist_KT1.tex}  
  \caption{Differential distributions with respect to the transverse
    momentum of the leading jet $\ktl$ at the LHC with CM energies 7~\TeV,
    8~\TeV, and 14~\TeV.
    Left: absolute predictions; right: relative contributions $\delta$.}
  \label{fig:delkt}
\end{figure}
The differential distributions in $\ktl$ shown in Fig.~\ref{fig:delkt}
cover the range up to $\ktl=1.5~\TeV$ for $\sqrt{s}=7~\TeV,\,8~\TeV$, and
up to $\ktl=3~\TeV$ for $\sqrt{s}=14~\TeV$ and further underline the
observations made above. 
The weak corrections display the expected behaviour from the Sudakov
logarithms with corrections reaching up to $-6\%$ at $\ktl=1.5~\TeV$
for $\sqrt{s}=~7~\TeV,\,8~\TeV$, and $-12\%$ at $\ktl=3~\TeV$
for $\sqrt{s}=~14~\TeV$.
On the other hand, the tree-level corrections $\deltree$
increase with higher $\ktl$ and reach
approximately $16\%$ at $\ktl=1.5~\TeV$
for $\sqrt{s}=~7~\TeV,\,8~\TeV$, and $20\%$ at $\ktl=3~\TeV$
for $\sqrt{s}=~14~\TeV$, resulting in
significant cancellations in the sum $\delsum$.

% \begin{figure}
%   \centering
%   \scriptsize
%   \input{plots/cdj7dist_KT1_DBLE.tex}\\
%   \input{plots/cdj7dist_KT1_DBLE_delta.tex}
%   \caption{Double-differential distribution with respect to the
%     transverse momentum of the leading jet $\ktl$ and $\ystar$ at the
%     LHC with a CM energy of 7~\TeV.}
%   \label{fig:dbledelkt7}
% \end{figure}
\begin{figure}
  \centering
  \scriptsize
  \input{plots/cdj8dist_KT1_DBLE.tex}\\
  \input{plots/cdj8dist_KT1_DBLE_delta.tex}
  \caption{Double-differential distribution with respect to the
    transverse momentum of the leading jet $\ktl$ and $\ystar$ at the
    LHC with a CM energy of 8~\TeV. In the absolute prediction (uppermost plot)
    the cross section is divided by the bin width in $\ystar$.}
  \label{fig:dbledelkt8}
\end{figure}
\begin{figure}
  \centering
  \scriptsize
  \input{plots/cdj14dist_KT1_DBLE.tex}\\
  \input{plots/cdj14dist_KT1_DBLE_delta.tex}
  \caption{Same as in Fig.~\ref{fig:dbledelkt8}, but for a CM energy
    of 14~\TeV.}
  \label{fig:dbledelkt14}
\end{figure}
By introducing a further binning in $\ystar$ we obtain the double
differential distributions shown in
Figs.~\ref{fig:dbledelkt8} and \ref{fig:dbledelkt14}.
At higher values of the transverse momentum, the production of the
jets is required to be more and more central in the partonic CM frame,
leading to the observed rapid decrease in the cross section for the
bins with higher values of $\ystar$. 
In contrast to the $\mjj$ distribution, the bin with the smallest value
for $\ystar$ is the most dominant in the high-$\ktl$ tail.
Moreover, both the tree-level EW corrections $\deltree$ and the one-loop
weak radiative corrections $\delloop$ are only slightly affected
by the $\ystar$ binning, while there is a significant dependence
in the $\mjj$ distribution discussed in the previous section.

\begin{figure}
  \centering
  \scriptsize
  \input{plots/cdj7dist_KT2.tex}\\
  \input{plots/cdj8dist_KT2.tex}\\
  \input{plots/cdj14dist_KT2.tex}  
  \caption{Differential distributions with respect to the transverse
    momentum of the subleading jet $\ktsl$ at the LHC with CM energies 7~\TeV,
    8~\TeV, and 14~\TeV.
    Left: absolute predictions; right: relative contributions $\delta$.}
  \label{fig:delkts}
\end{figure}
% \begin{figure}
%   \centering
%   \scriptsize
%   \input{plots/cdj7dist_KT2_DBLE.tex}\\
%   \input{plots/cdj7dist_KT2_DBLE_delta.tex}
%   \caption{Double-differential distribution with respect to the
%     transverse momentum of the subleading jet $\ktsl$ and $\ystar$ at the
%     LHC with a CM energy of 7~\TeV.}
%   \label{fig:dbledelkts7}
% \end{figure}
\begin{figure}
  \centering
  \scriptsize
  \input{plots/cdj8dist_KT2_DBLE.tex}\\
  \input{plots/cdj8dist_KT2_DBLE_delta.tex}
  \caption{Double-differential distribution with respect to the
    transverse momentum of the subleading jet $\ktsl$ and $\ystar$ at the
    LHC with a CM energy of 8~\TeV.
    In the absolute prediction (uppermost plot)
    the cross section is divided by the bin width in $\ystar$.}
  \label{fig:dbledelkts8}
\end{figure}
\begin{figure}
  \centering
  \scriptsize
  \input{plots/cdj14dist_KT2_DBLE.tex}\\
  \input{plots/cdj14dist_KT2_DBLE_delta.tex}
  \caption{Same as in Fig.~\ref{fig:dbledelkts8}, but for a CM energy
    of 14~\TeV.}
  \label{fig:dbledelkts14}
\end{figure}
Additionally, we present the corresponding transverse-momentum
distributions with respect
to the subleading jet in
Figs.~\ref{fig:delkts} and \ref{fig:dbledelkts14}. 
Recall that leading and subleading jets have the same transverse momenta
($\ktl=\ktsl$) in all $2\to2$ particle configurations, \ie that only
the real emission corrections to the four-quark channels show
a different behaviour here. In particular, $\deltree$ remains the same
when going from the leading to the subleading jet.
On the other hand, the weak loop corrections
$\delloop$ turn out to be more pronounced for small $\ystar$.
This is due to the fact that subleading jets 
fill bins with smaller $\kt$ in the spectra which rapidly decrease
with higher $\kt$'s. 
Since the positive real-emission contribution on the $\kt$ axis 
of the subleading jets is, thus, shifted to the left as compared to 
the leading jet, the sum of negative virtual and
positive real corrections for a fixed bin is somewhat shifted to more
negative values for the subleading jet. This effect is largest
for the smallest $\ystar$ where the cross section is largest, because
the real emission has a particular tendency to reduce maxima in distributions.

\subsubsection{Dijet production at the Tevatron}
\label{sec:tevatron}

In the following, we present the results for dijet production at the
Tevatron, \ie for a $\Pp\Pap$ initial state and a CM energy of
$\sqrt{s}=1.96~\TeV$.

\begin{figure}
  \centering
  \scriptsize
  \input{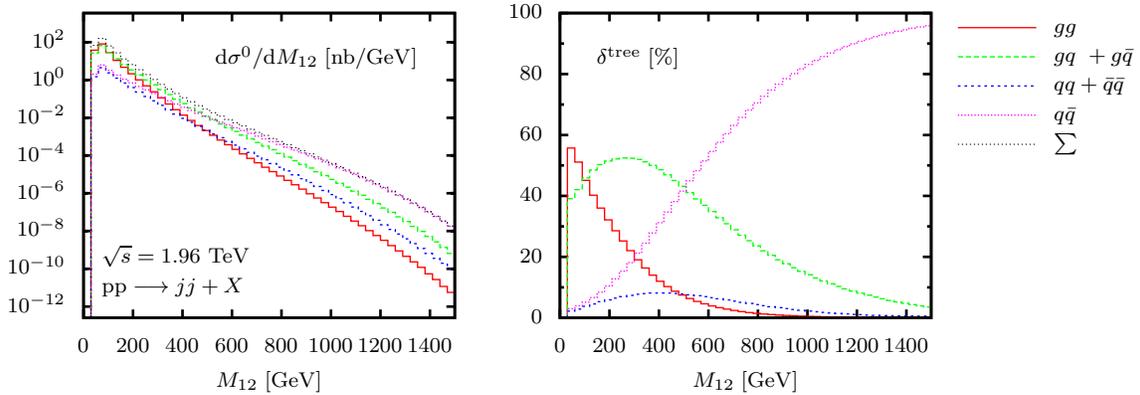}
  \caption{The LO contributions to the dijet invariant mass $\mjj$
    distribution from the different initial-state parton combinations
    at the Tevatron. 
    Left: absolute predictions; right: relative contributions
    $\delta^\text{tree}$.}
  \label{fig:tevmjjpids}
\end{figure}

Figure~\ref{fig:tevmjjpids} shows the various contributions of the
different partonic channels at LO contributing to dijet production
at the Tevatron for the invariant-mass distribution of the two jets.
Since the Tevatron is a $\Pp\Pap$ collider, with valence quark--antiquark
pairs in the initial state, there is a strong $q\bar q$ dominance at 
large values of the invariant mass $\mjj$ (several hundred GeV), which requires large
scattering energies and thus large momentum fraction $x$ of the partons.
At the moderate values $\mjj\lsim500~\GeV$, there is still some dominance of
channels with gluons in the initial state, with even the largest contribution from 
$\Pg\Pg$ scattering for very low $\mjj$, because the gluon PDF has the strongest
rise at small $x$.

\begin{figure}
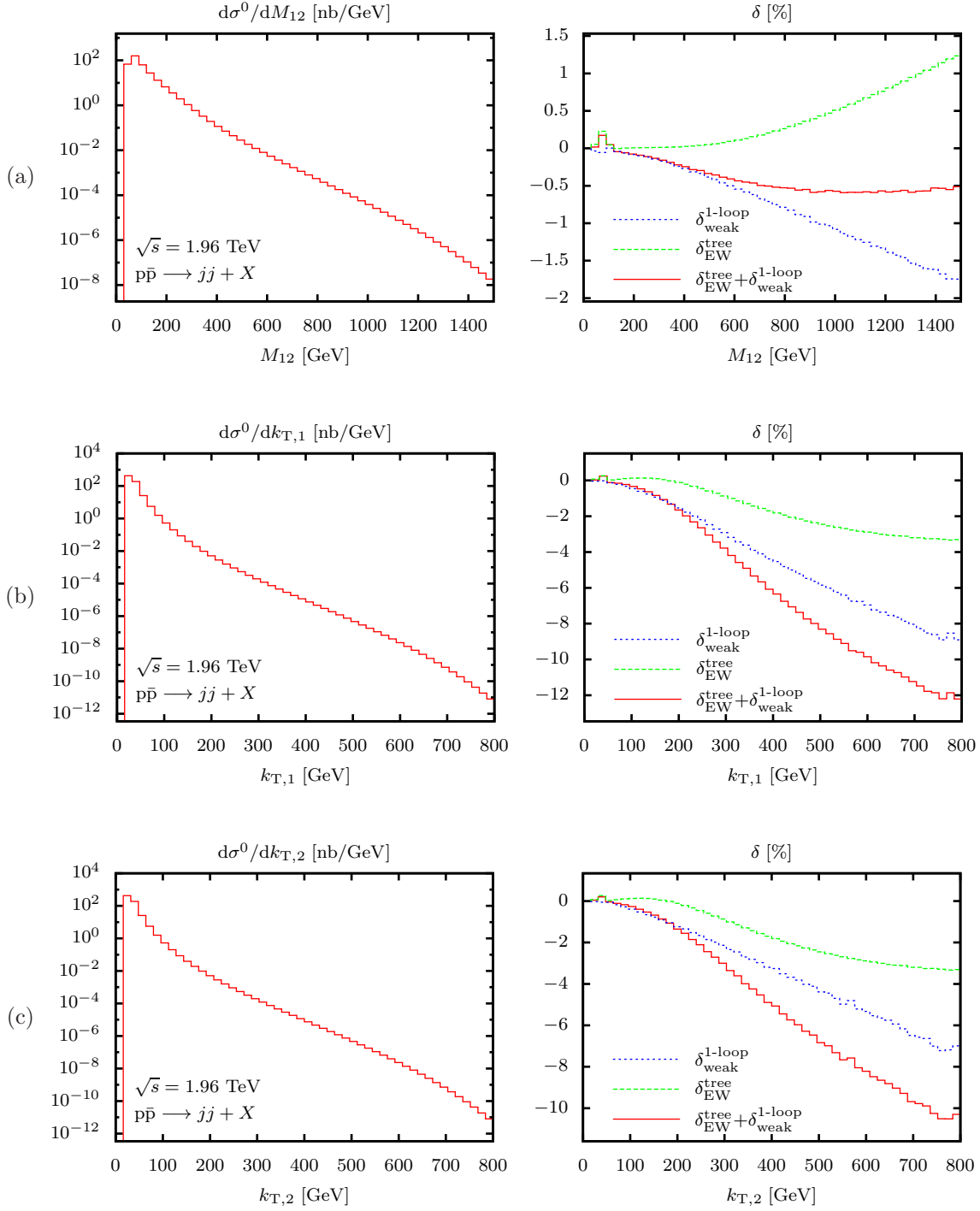

  \centering
  \scriptsize
  \raisebox{3cm}{\footnotesize(a)}~\input{plots/cdj2dist_MJJ.tex}\\
  \raisebox{3cm}{\footnotesize(b)}~\input{plots/cdj2dist_KT1.tex}\\
  \raisebox{3cm}{\footnotesize(c)}~\input{plots/cdj2dist_KT2.tex}
  \caption{Differential distributions with respect to the 
    dijet invariant mass $\mjj$~(a), the transverse momentum of the
    leading jet $\ktl$~(b) and the subleading jet $\ktsl$~(c) at the
    Tevatron for a CM energy of 1.96~\TeV.
    Left: absolute predictions; right: relative contributions $\delta$.}
  \label{fig:tevdist}
\end{figure}

The differential distributions with respect to the dijet invariant
mass $\mjj$, the transverse momentum of the leading jet $\ktl$ and the
subleading jet $\ktsl$ are shown in Fig.~\ref{fig:tevdist}(a--c),
respectively. 
In accordance with the observations made for the LHC, the
weak corrections $\delloop$ are much smaller for $\mjj$-based
observables as compared to those based on the transverse momenta of the
jets. 
We further observe that the LO EW contributions $\deltree$ in the
$\mjj$ distribution is similar in magnitude and opposite in sign
as compared to $\delloop$, similarly to the case of the LHC.
For the transverse-momentum distributions, however, $\deltree$ becomes
negative for higher $\kti$,  further increasing the corrections in
$\delsum$, which reach around $-12\%$ for $\kti=800~\GeV$. 
It is interesting to note that the radiative corrections $\delloop$
are similar for both, the leading and subleading jet $\kt$,
which is different from the behaviour we observed at the LHC,
where the corrections to $\ktsl$ are significantly larger.
To understand this, recall that the difference in the corrections
for leading and subleading jets is merely due to real emission
corrections in the four-quark channels. These are the real QCD corrections 
to the interference of weak and QCD tree diagrams. Generically these
interferences, which are part of $\deltree$ in LO, are much smaller
at the Tevatron than at the LHC, as is obvious from $\deltree$ in 
Fig.~\ref{fig:tevdist}. Moreover, it is interesting
that the corrections are even slightly larger for the leading jet
than for the subleading one, because the real QCD corrections to the
weak--QCD interference are negative at high energy scales, 
as also observed for $\deltree$ before.

\begin{figure}
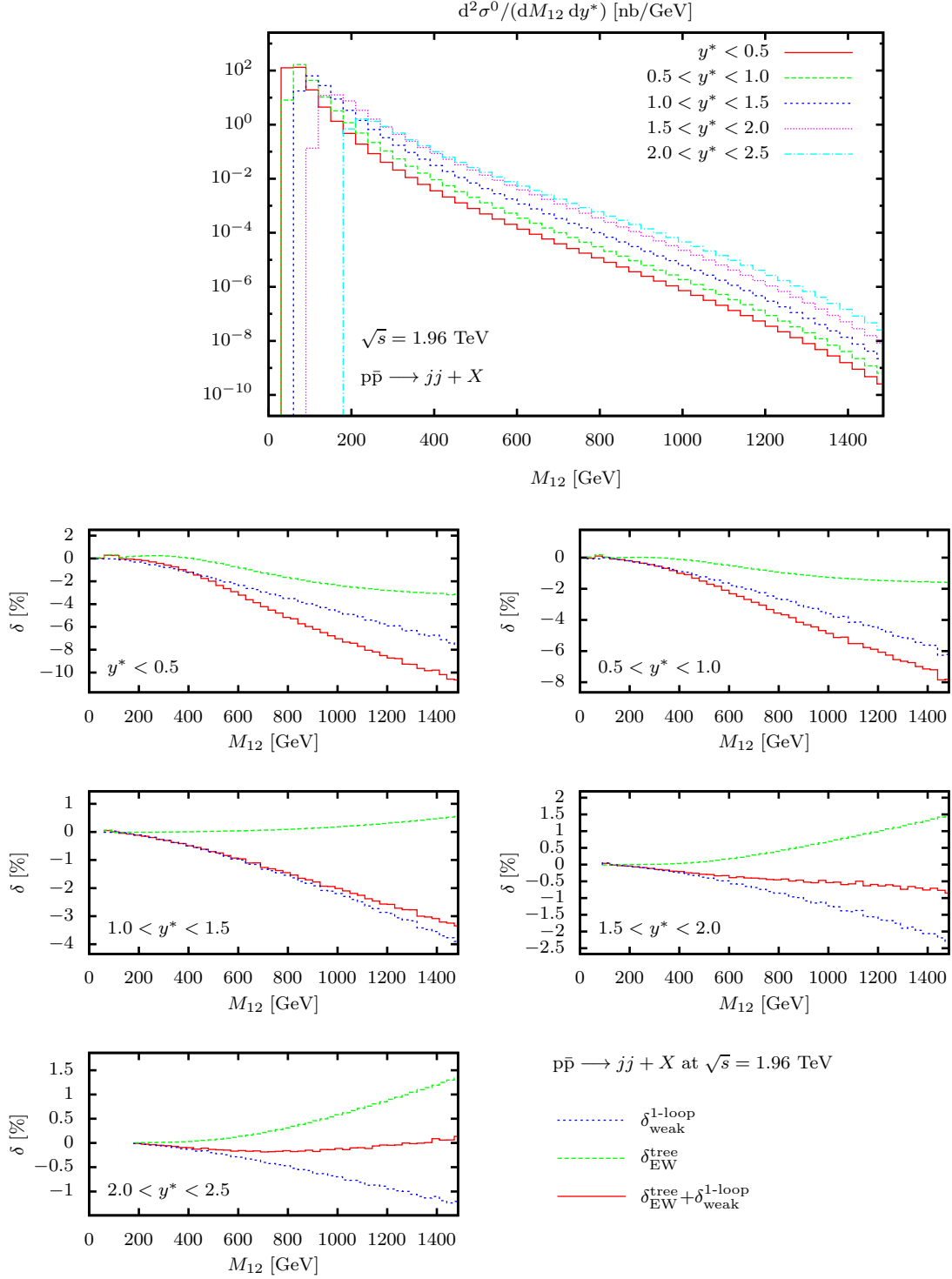

  \centering
  \scriptsize
  \input{plots/cdj2dist_MJJ_DBLE.tex}\\
  \input{plots/cdj2dist_MJJ_DBLE_delta.tex}
  \caption{Double-differential distribution with respect to the
    dijet invariant mass $\mjj$ and $\ystar$ at the
    Tevatron.
    In the absolute prediction (uppermost plot)
    the cross section is divided by the bin width in $\ystar$.}
  \label{fig:tevdblemjj}
\end{figure}

The $\mjj$ and $\ktl$ distributions with a further binning in $\ystar$ are shown
in Figs.~\ref{fig:tevdblemjj} and \ref{fig:tevdblektl}.
\begin{figure}
  \centering
  \scriptsize
  \input{plots/cdj2dist_KT1_DBLE.tex}\\
  \input{plots/cdj2dist_KT1_DBLE_delta.tex}
  \caption{Double-differential distribution with respect to the
    transverse momentum of the leading jet $\ktl$ and $\ystar$ at the
    Tevatron.
    In the absolute prediction (uppermost plot)
    the cross section is divided by the bin width in $\ystar$.}
  \label{fig:tevdblektl}
\end{figure}
%\begin{figure}
%  \centering
%  \scriptsize
%  \input{plots/cdj2dist_KT2_DBLE.tex}\\
%  \input{plots/cdj2dist_KT2_DBLE_delta.tex}
%  \caption{Double-differential distribution with respect to the
%    transverse momentum of the subleading jet $\ktsl$ and $\ystar$ at the
%    Tevatron.
%    In the absolute prediction (uppermost plot)
%    the cross section is divided by the bin width in $\ystar$.}
%  \label{fig:tevdblektsl}
%\end{figure}

\subsubsection{Comparison to other work}
\label{sec:comp}

Preliminary results for the weak radiative corrections to dijet
production at the LHC have been presented by A.~Scharf in the
proceedings contribution~\cite{Scharf2009}, where the contributions from
external bottom quarks were 
not considered as part of dijet production, but
discussed separately.
For comparison, we here
adopt the calculational setup of Ref.~\cite{Scharf2009} for the LHC,
\begin{align}
  \sqrt{s} &= 14~\TeV, &
  \ktl^\cut = \ktsl^\cut &= 50~\GeV, \nnnl
  y^\cut &: \text{none}, &
  \mu_\rF=\mu_\rR = 2k_{\rT}^\cut &= 100~\GeV, \nnnl
  \alphas &= 0.1, & \alpha &= 1/128, \nnnl
  \label{eq:3}
  M_\PW &= 80.425~\GeV, & M_\PZ &= 91.1876~\GeV , \nnnl
  \text{PDF set} &: \text{CTEQ6L},
\end{align}
which is partially inferred from Ref.~\cite{Kuhn2010}.
According to the author of Ref.~\cite{Scharf2009}, further
details on the jet algorithm and the precise treatment of the 
W/Z resonances are not available anymore,
but those loose ends should only play a minor role.
\begin{figure}
  \centering
  \footnotesize
  \input{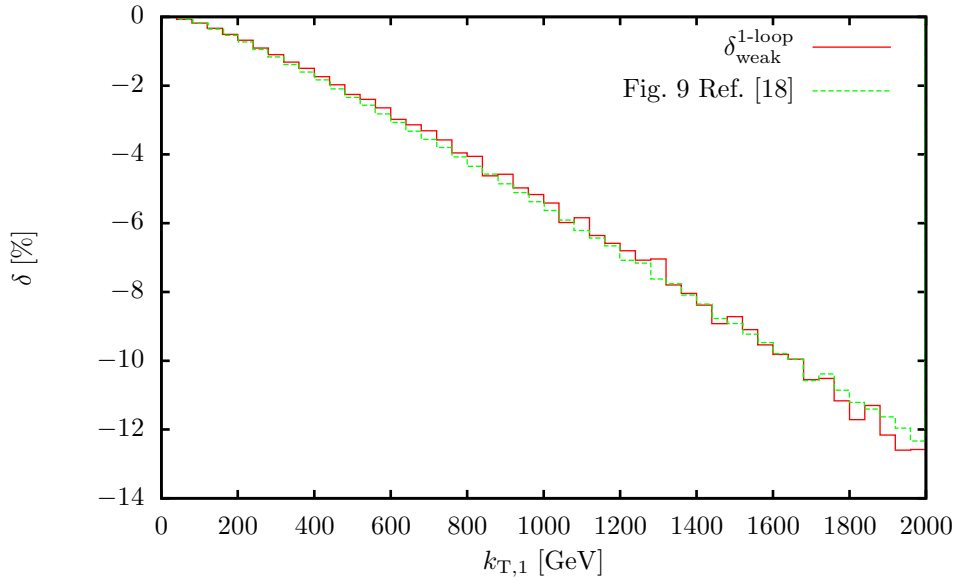}
  \caption{Comparison of the weak one-loop correction $\delloop$ to the 
   transverse-momentum spectrum of the leading jet as obtained from our
   calculation to the result of Ref.~\cite{Scharf2009}.} 
  \label{fig:scharf}
\end{figure}
This is confirmed in 
Fig.~\ref{fig:scharf} which shows our result on the
corrections to the transverse-momentum spectrum of the leading jet
in comparison to the one shown
to Fig.~9 of Ref.~\cite{Scharf2009}.
The two results show good agreement over the considered $\ktl$
range. 

We have also tried to perform a tuned comparison to the results of
Ref.~\cite{Moretti2006c}, but have not found agreement.%
\footnote{To find the source of discrepancy seems to require a careful comparison
of individual components of the calculation. Since the correctness of our
results is backed by our two calculations and the comparison to
Ref.~\cite{Scharf2009}, we do not see a reason to await the outcome
of this procedure before publication.}

\section{Conclusions}
\label{sec:concl}

In this paper
we have presented the calculation of the most important electroweak
corrections to dijet production at the LHC and the Tevatron.
These corrections comprise electroweak contributions of
$\order{\alphas\alpha,\,\alpha^2}$ to the LO QCD prediction
as well as NLO corrections 
through the order $\alphas^2\alpha$.
Guided by the electroweak Sudakov-type logarithms induced by soft
or collinear W/Z exchange at high energies, we have restricted
ourselves to the calculation of the purely weak loop corrections in a
first step. 

For the integrated cross section with minimal cuts
we find that the weak corrections
are negligible, typically staying below the per-cent level, both at
the LHC and the Tevatron. 
However, the electroweak Sudakov logarithms 
affect the tails of kinematic distributions that are sensitive
to the high energy scales of the hard scattering process.
We have discussed this feature in some detail, considering
the distributions in the dijet
invariant mass ($\mjj$), and in the transverse momenta of the leading
($\ktl$) and the subleading ($\ktsl$) jets.

For the $\sqrt{s}=14~\TeV$ LHC setup, 
we observe weak loop corrections of $\order{\alphas^2\alpha}$
reaching up to $-12\%$ ($-16\%$) for 
a transverse momentum of $\kt=3~\TeV$ of the leading (subleading) jet, whereas
the dijet invariant-mass distribution only receives weak
corrections up to $-6\%$ for $\mjj=6~\TeV$.
This difference is explained by the fact that observables based on
specific ranges in $\mjj$ are not dominated by the Sudakov regime
(large energies at fixed angles)
for large $\mjj$ values, but rather characterized by
the Regge (forward) regime.
The weak corrections to dijet production at the Tevatron show
similar features, though their size is smaller 
($-10\%$ for $\kt=800~\GeV$)
than for the LHC because of the smaller scattering energy.

The LO EW contributions of $\order{\alphas\alpha,\,\alpha^2}$ 
turn out to be of the same order of magnitude as the weak loop
corrections.
At the LHC, these two types of corrections partially cancel, but
the degree of this cancellation depends on the chosen observable,
setup, and cuts, so that the full calculation is necessary in
order to correctly include the considered electroweak effects.
At the Tevatron the LO electroweak corrections are somewhat smaller
than the weak loop effects, but of the same sign
at high transverse jet momenta,
thus somewhat enhancing the electroweak effects 
to $-12\%$ at $\kt=800~\GeV$.

The electroweak corrections considered in this paper are supposed to be 
the by far dominant electroweak effects in dijet production at hadron
colliders. Their numerical impact of $10{-}20\%$ in the TeV
range is not negligible and will certainly play a significant role
once the NNLO QCD corrections are known.
In contrast to the weak corrections, which get dominated
by the large Sudakov logarithms at high energies,
the so-far-neglected photonic loop corrections do not
receive particular enhancements over their parametric suppression
by the electromagnetic coupling $\alpha$. The calculation of these
effects, which are expected to stay at the few-per-cent level,
are left to the future.

\section*{Acknowledgements}

A.H. is supported by the German Research Foundation (DFG) via grant DI
784/2-1.

\input{refs.bbl}
\end{document}

%% file: tables/cdj7totmjj.tex
% generated by combtot.sh on Thu Sep 20 18:11:03 CEST 2012 
{ \renewcommand{\arraystretch}{1.2} $$ \begin{array}{c|c c c c c c c }
\multicolumn{8}{c}{\proc\;\mbox{at}\;\sqrt{s}=7~\TeV}\\\hline
\mjj/\GeV
&	50-\infty
&	100-\infty
&	200-\infty
&	500-\infty
&	1000-\infty
&	2000-\infty
&	3000-\infty
\\\hline
\sigtree/\nb
%	7.859993e+04	1.808384e+00
&	78600(2)
%	2.549555e+04	1.099033e+00
&	25496(1)
%	3.879165e+03	2.607711e-01
&	3879.2(3)
%	8.080748e+01	1.930958e-03
&	80.807(2)
%	2.493136e+00	5.351732e-05
&	2.49314(5)
%	3.304867e-02	9.292004e-07
&	33.0487(9)\cdot10^{-3}
%	8.908274e-04	1.446854e-08
&	890.83(1)\cdot10^{-6}
\\
\sigqcd/\nb
%	7.856103e+04	1.633454e+00
&	78561(2)
%	2.548981e+04	1.080679e+00
&	25490(1)
%	3.878069e+03	9.556756e-02
&	3878.07(10)
%	8.064953e+01	2.560981e-03
&	80.650(3)
%	2.475644e+00	1.610100e-04
&	2.4756(2)
%	3.248258e-02	8.191619e-07
&	32.4826(8)\cdot10^{-3}
%	8.706889e-04	2.310375e-08
&	870.69(2)\cdot10^{-6}
\\\hline
\delloop/\%
%	-2.394846e-02	-2.226643e-04
&	-0.02
%	-3.325776e-02	-1.121766e-04
&	-0.03
%	-8.824035e-02	-1.176634e-04
&	-0.09
%	-3.905388e-01	-5.729318e-04
&	-0.39
%	-1.085358e+00	-1.453869e-03
&	-1.09
%	-2.478331e+00	-3.049014e-03
&	-2.48
%	-3.600623e+00	-4.482216e-03
&	-3.60
\\
\deltree/\%
%	4.789418e-02	1.104540e-06
&	 0.05
%	1.371953e-02	6.780384e-07
&	 0.01
%	3.528282e-02	1.377095e-06
&	 0.04
%	1.973011e-01	6.882679e-06
&	 0.20
%	7.025505e-01	4.751120e-05
&	 0.70
%	1.745646e+00	7.917098e-05
&	 1.75
%	2.310008e+00	7.012573e-05
&	 2.31
\\
\delsumabbr/\%
%	2.394572e-02	2.226648e-04
&	 0.02
%	-1.953824e-02	-1.121713e-04
&	-0.02
%	-5.295753e-02	-1.176554e-04
&	-0.05
%	-1.932377e-01	-5.728375e-04
&	-0.19
%	-3.828074e-01	-1.452427e-03
&	-0.38
%	-7.326855e-01	-3.049139e-03
&	-0.73
%	-1.290616e+00	-4.481458e-03
&	-1.29
\\\hline
\sigloop/\nb
%	7.858112e+04	1.816825e+00
&	78581(2)
%	2.548707e+04	1.099405e+00
&	25487(1)
%	3.875743e+03	2.608110e-01
&	3875.7(3)
%	8.049251e+01	1.985448e-03
&	80.493(2)
%	2.466266e+00	6.447107e-05
&	2.46627(6)
%	3.224364e-02	1.357899e-06
&	32.244(1)\cdot10^{-3}
%	8.594772e-04	4.161355e-08
&	859.48(4)\cdot10^{-6}
\end{array} $$ }

%% file: tables/cdj8totmjj.tex
% generated by combtot.sh on Thu Sep 20 18:11:06 CEST 2012 
{ \renewcommand{\arraystretch}{1.2} $$ \begin{array}{c|c c c c c c c }
\multicolumn{8}{c}{\proc\;\mbox{at}\;\sqrt{s}=8~\TeV}\\\hline
\mjj/\GeV
&	50-\infty
&	100-\infty
&	200-\infty
&	500-\infty
&	1000-\infty
&	2000-\infty
&	3000-\infty
\\\hline
\sigtree/\nb
%	9.468273e+04	2.083611e+00
&	94683(2)
%	3.120075e+04	1.499535e+00
&	31201(1)
%	4.868225e+03	2.979893e-01
&	4868.2(3)
%	1.083311e+02	7.489544e-03
&	108.331(7)
%	3.656845e+00	7.301573e-05
&	3.65684(7)
%	6.149779e-02	1.147689e-06
&	61.498(1)\cdot10^{-3}
%	2.454754e-03	6.101201e-08
&	2.45475(6)\cdot10^{-3}
\\
\sigqcd/\nb
%	9.463849e+04	2.020216e+00
&	94638(2)
%	3.119664e+04	1.812092e+00
&	31197(2)
%	4.867071e+03	1.415720e-01
&	4867.1(1)
%	1.081519e+02	3.004203e-03
&	108.152(3)
%	3.634375e+00	1.326948e-04
&	3.6344(1)
%	6.052372e-02	9.346872e-07
&	60.5237(9)\cdot10^{-3}
%	2.400560e-03	5.025671e-08
&	2.40056(5)\cdot10^{-3}
\\\hline
\delloop/\%
%	-2.252160e-02	-4.818815e-04
&	-0.02
%	-3.214594e-02	-1.127511e-04
&	-0.03
%	-8.493050e-02	-1.387814e-04
&	-0.08
%	-3.740709e-01	-4.885379e-04
&	-0.37
%	-1.051106e+00	-1.691828e-03
&	-1.05
%	-2.442735e+00	-2.696461e-03
&	-2.44
%	-3.622894e+00	-4.701804e-03
&	-3.62
\\
\deltree/\%
%	4.416566e-02	1.043858e-06
&	 0.04
%	1.233451e-02	7.587365e-07
&	 0.01
%	3.135981e-02	1.478877e-06
&	 0.03
%	1.710997e-01	5.126553e-06
&	 0.17
%	6.134414e-01	2.410442e-05
&	 0.61
%	1.610041e+00	3.132815e-05
&	 1.61
%	2.252644e+00	5.421649e-05
&	 2.25
\\
\delsumabbr/\%
%	2.164406e-02	4.818817e-04
&	 0.02
%	-1.981142e-02	-1.127418e-04
&	-0.02
%	-5.357070e-02	-1.387730e-04
&	-0.05
%	-2.029712e-01	-4.884637e-04
&	-0.20
%	-4.376645e-01	-1.691491e-03
&	-0.44
%	-8.326945e-01	-2.696295e-03
&	-0.83
%	-1.370251e+00	-4.701356e-03
&	-1.37
\\\hline
\sigloop/\nb
%	9.466142e+04	2.132935e+00
&	94661(2)
%	3.119072e+04	1.499947e+00
&	31191(1)
%	4.864091e+03	2.980658e-01
&	4864.1(3)
%	1.079265e+02	7.508150e-03
&	107.926(8)
%	3.618644e+00	9.544657e-05
&	3.61864(10)
%	6.001936e-02	1.995016e-06
&	60.019(2)\cdot10^{-3}
%	2.367784e-03	1.282915e-07
&	2.3678(1)\cdot10^{-3}
\end{array} $$ }

%% file: tables/cdj14totmjj.tex
% generated by combtot.sh on Thu Sep 20 18:11:14 CEST 2012 
{ \renewcommand{\arraystretch}{1.2} $$ \begin{array}{c|c c c c c c c }
\multicolumn{8}{c}{\proc\;\mbox{at}\;\sqrt{s}=14~\TeV}\\\hline
\mjj/\GeV
&	50-\infty
&	100-\infty
&	200-\infty
&	500-\infty
&	1000-\infty
&	2000-\infty
&	5000-\infty
\\\hline
\sigtree/\nb
%	1.984830e+05	4.458808e+00
&	198483(4)
%	6.933466e+04	3.420601e+00
&	69335(3)
%	1.185817e+04	8.379181e-01
&	11858.2(8)
%	3.341710e+02	1.227662e-02
&	334.17(1)
%	1.494351e+01	4.094900e-04
&	14.9435(4)
%	4.566424e-01	1.377807e-05
&	456.64(1)\cdot10^{-3}
%	9.547117e-04	1.913948e-08
&	954.71(2)\cdot10^{-6}
\\
\sigqcd/\nb
%	1.984097e+05	4.636796e+00
&	198410(5)
%	6.932899e+04	1.727227e+00
&	69329(2)
%	1.185595e+04	9.757964e-01
&	11856.0(10)
%	3.338491e+02	7.516568e-03
&	333.849(8)
%	1.489376e+01	3.446969e-04
&	14.8938(3)
%	4.521157e-01	1.646049e-05
&	452.12(2)\cdot10^{-3}
%	9.309043e-04	2.590525e-08
&	930.90(3)\cdot10^{-6}
\\\hline
\delloop/\%
%	-1.985895e-02	-1.862435e-04
&	-0.02
%	-2.853568e-02	-1.184565e-04
&	-0.03
%	-7.342120e-02	-1.290252e-04
&	-0.07
%	-3.135106e-01	-4.807673e-04
&	-0.31
%	-8.848634e-01	-9.705818e-04
&	-0.88
%	-2.200756e+00	-3.050311e-03
&	-2.20
%	-5.534545e+00	-7.409396e-03
&	-5.53
\\
\deltree/\%
%	3.295839e-02	8.421748e-07
&	 0.03
%	8.552843e-03	3.037880e-07
&	 0.01
%	2.056071e-02	1.720687e-06
&	 0.02
%	9.830432e-02	3.058056e-06
&	 0.10
%	3.376597e-01	9.119489e-06
&	 0.34
%	1.002019e+00	4.284325e-05
&	 1.00
%	2.557644e+00	8.210785e-05
&	 2.56
\\
\delsumabbr/\%
%	1.309944e-02	1.862435e-04
&	 0.01
%	-1.998284e-02	-1.184556e-04
&	-0.02
%	-5.286050e-02	-1.289574e-04
&	-0.05
%	-2.152063e-01	-4.807445e-04
&	-0.22
%	-5.472038e-01	-9.704598e-04
&	-0.55
%	-1.198738e+00	-3.049653e-03
&	-1.20
%	-2.976901e+00	-7.408371e-03
&	-2.98
\\\hline
\sigloop/\nb
%	1.984436e+05	4.474094e+00
&	198444(4)
%	6.931488e+04	3.421587e+00
&	69315(3)
%	1.184947e+04	8.380574e-01
&	11849.5(8)
%	3.331243e+02	1.238107e-02
&	333.12(1)
%	1.481172e+01	4.342456e-04
&	14.8117(4)
%	4.466924e-01	1.949087e-05
&	446.69(2)\cdot10^{-3}
%	9.031904e-04	7.156626e-08
&	903.19(7)\cdot10^{-6}
\end{array} $$ }

%% file: tables/cdj7totkt.tex
% generated by combtot.sh on Thu Sep 20 18:11:20 CEST 2012 
{ \renewcommand{\arraystretch}{1.2} $$ \begin{array}{c|c c c c c c c }
\multicolumn{8}{c}{\proc\;\mbox{at}\;\sqrt{s}=7~\TeV}\\\hline
\ktl/\GeV
&	25-\infty
&	50-\infty
&	100-\infty
&	200-\infty
&	500-\infty
&	1000-\infty
&	1500-\infty
\\\hline
\sigtree/\nb
%	7.859993e+04	1.808384e+00
&	78600(2)
%	5.417059e+03	1.313074e-01
&	5417.1(1)
%	2.912053e+02	8.901117e-03
&	291.205(9)
%	1.117648e+01	2.217915e-04
&	11.1765(2)
%	6.369731e-02	1.617000e-06
&	63.697(2)\cdot10^{-3}
%	3.968433e-04	8.093045e-09
&	396.843(8)\cdot10^{-6}
%	5.974615e-06	9.784728e-11
&	5.9746(10)\cdot10^{-6}
\\
\sigqcd/\nb
%	7.856103e+04	1.633454e+00
&	78561(2)
%	5.413752e+03	1.797100e-01
&	5413.8(2)
%	2.905168e+02	8.892367e-03
&	290.517(9)
%	1.108727e+01	3.175109e-04
&	11.0873(3)
%	6.125084e-02	1.886629e-06
&	61.251(2)\cdot10^{-3}
%	3.534723e-04	1.267268e-08
&	353.47(1)\cdot10^{-6}
%	4.961066e-06	8.820517e-11
&	4.9611(9)\cdot10^{-6}
\\\hline
\delloop/\%
%	-2.383698e-02	-1.896046e-04
&	-0.02
%	-9.566268e-02	-2.171252e-04
&	-0.10
%	-3.403619e-01	-6.420042e-04
&	-0.34
%	-9.931880e-01	-1.503548e-03
&	-0.99
%	-2.955174e+00	-5.433405e-03
&	-2.96
%	-5.115857e+00	-1.071982e-02
&	-5.12
%	-6.129346e+00	-1.649374e-02
&	-6.13
\\
\deltree/\%
%	4.789418e-02	1.104540e-06
&	 0.05
%	5.546660e-02	2.054947e-06
&	 0.06
%	2.363424e-01	9.248372e-06
&	 0.24
%	8.000980e-01	2.749404e-05
&	 0.80
%	3.992655e+00	1.515372e-04
&	 3.99
%	1.227872e+01	4.591669e-04
&	12.28
%	2.043611e+01	4.330084e-04
&	20.44
\\
\delsumabbr/\%
%	2.405720e-02	1.896052e-04
&	 0.02
%	-4.019608e-02	-2.171080e-04
&	-0.04
%	-1.040194e-01	-6.419535e-04
&	-0.10
%	-1.930900e-01	-1.503366e-03
&	-0.19
%	1.037481e+00	5.433458e-03
&	 1.04
%	7.162859e+00	1.072212e-02
&	 7.16
%	1.430677e+01	1.649702e-02
&	14.31
\\\hline
\sigloop/\nb
%	7.858120e+04	1.814508e+00
&	78581(2)
%	5.411880e+03	1.318324e-01
&	5411.9(1)
%	2.902165e+02	9.094376e-03
&	290.216(9)
%	1.106636e+01	2.774369e-04
&	11.0664(3)
%	6.188724e-02	3.699623e-06
&	61.887(4)\cdot10^{-3}
%	3.787602e-04	3.874081e-08
&	378.76(4)\cdot10^{-6}
%	5.670534e-06	8.240771e-10
&	5.6705(8)\cdot10^{-6}
\end{array} $$ }

%% file: tables/cdj8totkt.tex
% generated by combtot.sh on Thu Sep 20 18:11:24 CEST 2012 
{ \renewcommand{\arraystretch}{1.2} $$ \begin{array}{c|c c c c c c c }
\multicolumn{8}{c}{\proc\;\mbox{at}\;\sqrt{s}=8~\TeV}\\\hline
\ktl/\GeV
&	25-\infty
&	50-\infty
&	100-\infty
&	200-\infty
&	500-\infty
&	1000-\infty
&	1500-\infty
\\\hline
\sigtree/\nb
%	9.468273e+04	2.083611e+00
&	94683(2)
%	6.762302e+03	2.276659e-01
&	6762.3(2)
%	3.806821e+02	1.317649e-02
&	380.68(1)
%	1.573869e+01	5.289871e-04
&	15.7387(5)
%	1.059649e-01	2.705167e-06
&	105.965(3)\cdot10^{-3}
%	8.778758e-04	2.341865e-08
&	877.88(2)\cdot10^{-6}
%	2.045375e-05	2.831862e-10
&	20.4538(3)\cdot10^{-6}
\\
\sigqcd/\nb
%	9.463849e+04	2.020216e+00
&	94638(2)
%	6.759007e+03	2.507612e-01
&	6759.0(3)
%	3.799042e+02	1.369266e-02
&	379.90(1)
%	1.563304e+01	5.762661e-04
&	15.6330(6)
%	1.026255e-01	2.207072e-06
&	102.626(2)\cdot10^{-3}
%	7.956965e-04	2.707332e-08
&	795.70(3)\cdot10^{-6}
%	1.734819e-05	2.327436e-10
&	17.3482(2)\cdot10^{-6}
\\\hline
\delloop/\%
%	-2.289815e-02	-2.091844e-04
&	-0.02
%	-9.150513e-02	-2.327129e-04
&	-0.09
%	-3.271198e-01	-5.867762e-04
&	-0.33
%	-9.699267e-01	-1.490660e-03
&	-0.97
%	-2.955645e+00	-3.984413e-03
&	-2.96
%	-5.351611e+00	-1.137990e-02
&	-5.35
%	-6.697670e+00	-1.513189e-02
&	-6.70
\\
\deltree/\%
%	4.416566e-02	1.043858e-06
&	 0.04
%	4.931797e-02	1.998161e-06
&	 0.05
%	2.051343e-01	8.552810e-06
&	 0.21
%	6.756037e-01	2.741998e-05
&	 0.68
%	3.257646e+00	1.002430e-04
&	 3.26
%	1.032924e+01	3.687409e-04
&	10.33
%	1.789953e+01	3.242529e-04
&	17.90
\\
\delsumabbr/\%
%	2.126751e-02	2.091848e-04
&	 0.02
%	-4.218716e-02	-2.326948e-04
&	-0.04
%	-1.219855e-01	-5.866900e-04
&	-0.12
%	-2.943230e-01	-1.490315e-03
&	-0.29
%	3.020018e-01	3.984556e-03
&	 0.30
%	4.977631e+00	1.138025e-02
&	 4.98
%	1.120186e+01	1.513394e-02
&	11.20
\\\hline
\sigloop/\nb
%	9.466106e+04	2.092995e+00
&	94661(2)
%	6.756117e+03	2.282085e-01
&	6756.1(2)
%	3.794394e+02	1.336365e-02
&	379.44(1)
%	1.558706e+01	5.780153e-04
&	15.5871(6)
%	1.029317e-01	4.902426e-06
&	102.932(5)\cdot10^{-3}
%	8.352932e-04	9.351755e-08
&	835.29(9)\cdot10^{-6}
%	1.929183e-05	2.640294e-09
&	19.292(3)\cdot10^{-6}
\end{array} $$ }

%% file: tables/cdj14totkt.tex
% generated by combtot.sh on Thu Sep 20 18:11:29 CEST 2012 
{ \renewcommand{\arraystretch}{1.2} $$ \begin{array}{c|c c c c c c c }
\multicolumn{8}{c}{\proc\;\mbox{at}\;\sqrt{s}=14~\TeV}\\\hline
\ktl/\GeV
&	25-\infty
&	50-\infty
&	100-\infty
&	200-\infty
&	500-\infty
&	1000-\infty
&	2500-\infty
\\\hline
\sigtree/\nb
%	1.984830e+05	4.458808e+00
&	198483(4)
%	1.619429e+04	4.690070e-01
&	16194.3(5)
%	1.074110e+03	2.558664e-02
&	1074.11(3)
%	5.640492e+01	1.547750e-03
&	56.405(2)
%	6.710980e-01	2.262001e-05
&	671.10(2)\cdot10^{-3}
%	1.203834e-02	3.938181e-07
&	12.0383(4)\cdot10^{-3}
%	8.850394e-06	1.911250e-10
&	8.8504(2)\cdot10^{-6}
\\
\sigqcd/\nb
%	1.984097e+05	4.636796e+00
&	198410(5)
%	1.618971e+04	5.008755e-01
&	16189.7(5)
%	1.072846e+03	3.927017e-02
&	1072.85(4)
%	5.620409e+01	1.490486e-03
&	56.204(1)
%	6.615178e-01	2.092846e-05
&	661.52(2)\cdot10^{-3}
%	1.150603e-02	1.890641e-07
&	11.5060(2)\cdot10^{-3}
%	7.482590e-06	1.576314e-10
&	7.4826(2)\cdot10^{-6}
\\\hline
\delloop/\%
%	-1.993599e-02	-1.770822e-04
&	-0.02
%	-7.845609e-02	-2.092094e-04
&	-0.08
%	-2.755799e-01	-4.079333e-04
&	-0.28
%	-8.352544e-01	-1.226370e-03
&	-0.84
%	-2.721458e+00	-3.511781e-03
&	-2.72
%	-5.482215e+00	-7.533470e-03
&	-5.48
%	-1.048829e+01	-1.791968e-02
&	-10.49
\\
\deltree/\%
%	3.295839e-02	8.421748e-07
&	 0.03
%	3.241775e-02	1.170712e-06
&	 0.03
%	1.209981e-01	5.103987e-06
&	 0.12
%	3.575713e-01	1.336055e-05
&	 0.36
%	1.443026e+00	5.345329e-05
&	 1.44
%	4.623521e+00	1.134857e-04
&	 4.62
%	1.827695e+01	5.582845e-04
&	18.28
\\
\delsumabbr/\%
%	1.302241e-02	1.770822e-04
&	 0.01
%	-4.603834e-02	-2.092010e-04
&	-0.05
%	-1.545818e-01	-4.078557e-04
&	-0.15
%	-4.776831e-01	-1.226272e-03
&	-0.48
%	-1.278432e+00	-3.511069e-03
&	-1.28
%	-8.586937e-01	-7.533416e-03
&	-0.86
%	7.788660e+00	1.792362e-02
&	 7.79
\\\hline
\sigloop/\nb
%	1.984434e+05	4.472629e+00
&	198443(4)
%	1.618159e+04	4.702283e-01
&	16181.6(5)
%	1.071153e+03	2.595801e-02
&	1071.15(3)
%	5.593547e+01	1.694246e-03
&	55.935(2)
%	6.530951e-01	3.241948e-05
&	653.10(3)\cdot10^{-3}
%	1.140755e-02	9.520154e-07
&	11.4076(10)\cdot10^{-3}
%	8.065599e-06	1.354308e-09
&	8.066(1)\cdot10^{-6}
\end{array} $$ }